\documentclass[acmtog]{acmart}

\usepackage{booktabs} 
\usepackage{xcolor}
\usepackage{float}

\usepackage{accents}
\usepackage{paralist}
\usepackage{listings}
\usepackage{sourcecodepro}
\usepackage[T1]{fontenc}
\usepackage{multicol}
\usepackage{subcaption}
\usepackage{stfloats}
\usepackage{soul}
\usepackage{multirow}
\usepackage[normalem]{ulem}
\usepackage[final]{pdfpages}

\renewcommand*\copyright{{\textcopyright}}

\newcommand{\SOUT}[1]{}
\newcommand{\changed}[2]{#2}
\newcommand{\revision}[1]{#1}

\definecolor{lightblue}{RGB}{240,245,255}
\definecolor{darkblue}{RGB}{40,40,85}

\usepackage[ruled]{algorithm2e} 

\SetAlFnt{\small}
\SetAlCapFnt{\small}
\SetAlCapNameFnt{\small}
\SetAlCapHSkip{0pt}
\usepackage{enumitem}
\setitemize{noitemsep,topsep=0pt,parsep=0pt,partopsep=0pt,leftmargin=*}

\newcommand{\reproduce}[1]{\textcolor{blue!20!black}{[Reproduce: \textbf{python3 #1}]}}
\newcommand{\reproducenvprof}[1]{\textcolor{blue!20!black}{[Reproduce: \textbf{nvprof --print-gpu-trace python3 #1}]}}

\begin{document}

\fancyfoot{}

\setcopyright{none}
\settopmatter{printacmref=false}
\renewcommand\footnotetextcopyrightpermission[1]{} 
\pagestyle{plain} 

\title{AsyncTaichi: On-the-fly Inter-kernel Optimizations for Imperative and Spatially Sparse Programming}


\author{Yuanming Hu}
\authornote{Both authors contributed equally to this work.}
\affiliation{
  \institution{Taichi Graphics \& MIT CSAIL}
}
\email{yuanming@taichi.graphics}
\author{Mingkuan Xu}
\authornotemark[1]
\affiliation{
  \institution{Taichi Graphics \& Tsinghua University}
}
\email{xmk17@mails.tsinghua.edu.cn}
\author{Ye Kuang}
\affiliation{
  \institution{Taichi Graphics}
}
\email{yekuang@taichi.graphics}
\author{Fr\'edo Durand}
\affiliation{
  \institution{MIT CSAIL}
}
\email{fredo@mit.edu}

%
%

%
%
\renewcommand{\shortauthors}{}

\begin{abstract}
\revision{Leveraging spatial sparsity has become a popular approach to accelerate 3D computer graphics applications. Spatially sparse data structures and efficient sparse kernels (such as parallel stencil operations on active voxels), are key to achieve high performance. Existing work focuses on improving performance within a single sparse computational kernel.
We show that, a system that looks beyond a single kernel, plus additional domain-specific sparse data structure analysis, opens up exciting new space for optimizing sparse computations.
Specifically, we propose a {\em domain-specific data-flow graph} model of {\em imperative and sparse} computation programs, which describes kernel relationships and enables easy analysis and optimization.
Combined with an asynchronous execution engine that exposes a wide window of kernels, the inter-kernel optimizer can then perform effective sparse computation optimizations, such as eliminating unnecessary voxel list generations and removing voxel activation checks. These domain-specific optimizations further make way for classical general-purpose optimizations that are originally challenging to directly apply to computations with sparse data structures.
{\em Without any computational code modification}, our new system leads to $4.02\times$ fewer kernel launches and $1.87\times$ speed up on our GPU benchmarks, including computations on Eulerian grids, Lagrangian particles, meshes, and automatic differentiation.}

\end{abstract}

\begin{teaserfigure}
\centering
\includegraphics[width=1.0\linewidth]{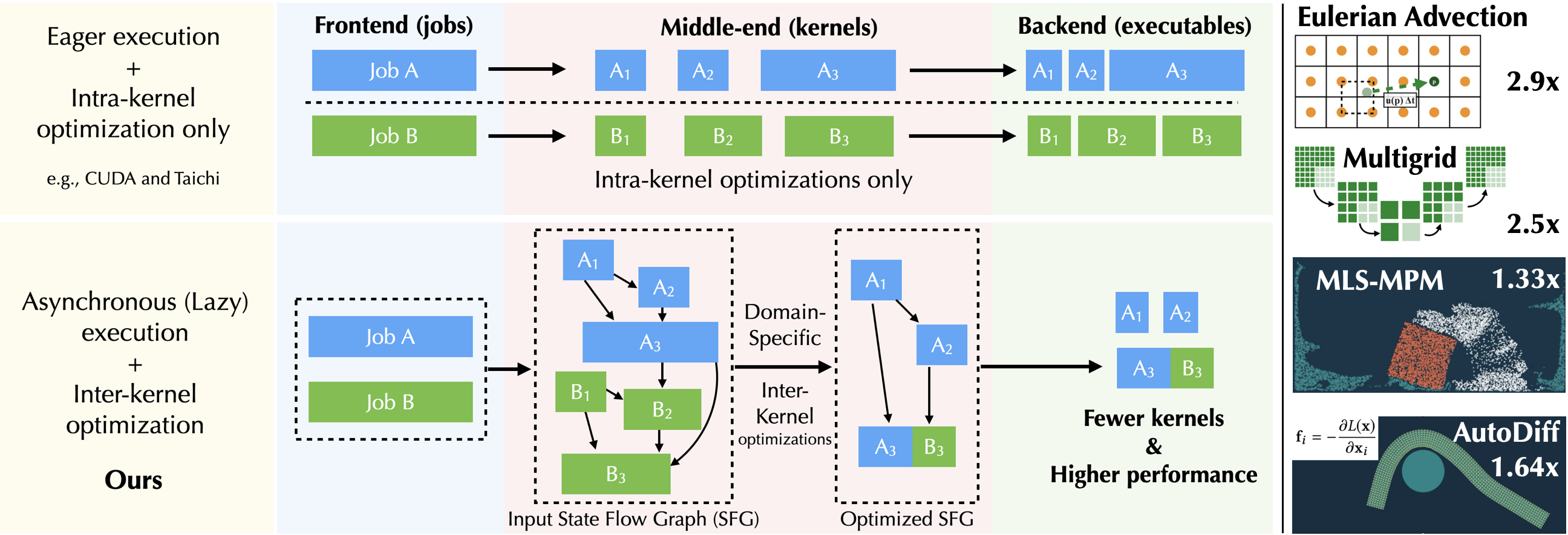}
\caption{\label{fig:teaser} \textbf{Top left:} In existing parallel imperative programming systems (such as CUDA and Taichi~\cite{hu2019taichi}), computational kernels are eagerly launched, leaving a tiny room for the optimizer to optimize {\em beyond a single kernel}. \textbf{Bottom left:} In this work, we accumulate kernels in an execution buffer, only flushing the execution queue when necessary. \revision{More importantly, we leverage a {\em domain-specific inter-kernel optimizer} to automatically conduct performance optimizations beyond a single kernel, especially for spatially sparse computation.} We dynamically build a tailored data-flow graph (``state-flow graph") of kernels for easy analysis, so that computation kernels can be optimized at a inter-kernel level just in time. \textbf{Right:} After a suite of domain-specific and general-purpose optimization passes including list generation removal, sparse data structure activation elimination, fusion, and dead store elimination, kernels are much better optimized compared to the reference intra-kernel optimization system. As a result, our inter-kernel optimized programs run $1.87\times$ faster on GPUs, {\em without the user modifying any of the computation code}. Our evaluation suite covers computations on Eulerian grids, Lagrangian particles, meshes, and automatic differentiation.}
\end{teaserfigure}

\maketitle
\thispagestyle{empty}

\section{Introduction}

\revision{Spatially sparse data strictures such as VDB~\cite{museth2013vdb} and SPGrid~\cite{setaluri2014spgrid} have been effective tools to improve performance and memory efficiency in various 3D computer graphics applications.
While existing work in sparse data structure libraries and compilers~\cite{wu2018fast, hoetzlein2016gvdb, gao2018gpu, hu2019taichi} have significantly improved sparse data structure performance within a single computational kernel, performance optimization opportunities remain abundant when considering multiple consecutive sparse computation kernel as a whole. For example, high-level knowledge of data structure sparsity patterns can often be extracted from the context around a sparse computation kernel. Such knowledge can help improve run-time performance of sparse data structure accesses, because the compiler may be able to infer whether the accessed voxel is active or not, thus saving sparsity checking overheads at run time.}

Optimizing across function and kernel boundaries is a well-developed technique in traditional ahead-of-time compilers (see, for example, gcc WHOPR~\cite{briggs2007whopr}), and functional array-based parallel JIT systems, especially deep learning systems such as TensorFlow~\cite{45381} and JAX~\cite{jax2018github}. However, two major challenges exist when applying the idea of inter-kernel optimization to spatially sparse computation programs.

\revision{Firstly, sparse data structures in graphics need to support {\em dynamic} topology. Operations on these data structures usually come with not only value modifications but also topology changes. Traditional compiler optimization tools for dense arrays with constant topology may not directly apply here.}

\revision{Secondly, graphics computations on sparse data structures are often {\em imperative} instead of functional, and side effects such as in-place modifications make imperative programming less ``pure'' and harder to optimize than functional programming. The reason why graphics programmers practically pick imperative rather than functional programming paradigm is two-fold: 1) sticking to imperative programming brings maximum compatibility with existing graphics algorithms, and 2) the tremendous amount of data in sparse computation implies programmers typically cannot afford to create a second copy of the data structures as in functional programming.}

\revision{As a result, classical optimizations, such as fusion, cannot be naturally used here, since the data structure topology might have changed between two kernels on sparse data structures.
Note that this is not an issue for programming systems with dense, immutable arrays. While tools for analyzing dense array programs are well established~\cite{knobe1998array}, their counterparts in {\em spatially sparse} array computations are largely underexploited.
Deep learning frameworks usually avoid some of the aforementioned challenges by adding constraints to the computation model, such as allowing only immutable data types to simplify code analysis. Although these constraints are often naturally satisfied in deep learning use cases, they may limit the programming flexibility in graphics (especially simulation) code.}

\revision{{\em Taichi}~\cite{hu2019taichi} is an imperative and spatially sparse domain-specific programming language that aims to achieve both high performance and high productivity. In this work, we concretely base our discussions on Taichi and strive to develop {\em an automatic inter-kernel optimization system for spatially sparse computation}.}

\paragraph{Design principles}
\revision{With productivity and performance in mind, our system is practically designed following the guidelines below:}
\begin{itemize}
    \item{\textbf{Transparent to users.}} We wish users can get the benefits of inter-kernel optimizations {\em for free}. No code modification should be needed in the computational kernels for users to leverage our optimizations.
    \revision{\item{\textbf{Just-in-time (JIT) compilation and optimization}.} JIT compilation is widely adopted by many Python-based computational frameworks (TensorFlow~\cite{45381}, PyTorch~\cite{NEURIPS2019_9015}, JAX~\cite{jax2018github}). It helps achieve a good balance between developer productivity and performance. However, JIT compilation alone usually cannot exploit the performance to its full extent. By accumulating the JIT-compiled kernel and building a {\em state-flow graph} on the fly, the Taichi runtime is able to uncover more opportunities for optimization.}
    \item{\textbf{Problems of all scales matter.}} Graphics applications cover a wide range of problem sizes. For example, a particle simulation may cover from $2$ thousand to $235$ million~\cite{hu2021quantaichi} particles. For small-scale tasks, {\em compilation} time may be the bottleneck; for large scale tasks, {\em computation} time is more important. Our asynchronous execution engine enables parallel compilation to reduce the JIT compilation delay.
\end{itemize}

\paragraph{Our solution} \revision{We propose a {\em domain-specific data-flow analysis model} of imperative and sparse programs, to analyze programs with partial and in-place updates and those with sparse data structures. ``States'' in our formulation refer not only to numerical values stored in data structures, but also to domain-specific data descriptions, such as the topology of the sparse data structures. We build a {\em state-flow graph (SFG)} on-the-fly consisting of pending kernels to depict kernel relationships. The rich expressiveness of SFGs allows us to conduct domain-specific inter-kernel optimizations, such as fusing kernels on {\em dynamic} sparse data structures, eliminate unnecessary voxel list generation tasks for parallel iterations on the sparse data structures, and accelerate the sparse data structure access.}

\revision{Imperative parallel programming systems often eagerly execute computational tasks, leaving little room to analyze and optimize beyond a single kernel. To enable the optimizer to see beyond a single kernel at a time, we built an {\em asynchronous execution engine} that maintains a window of kernels, leaving room for performance optimizations before launching the kernels. The execution engine also enables {\em parallel compilation}, which significantly reduces the JIT compilation time.}

\revision{We show that SFGs combined with the asynchronous engine open up space for inter-kernel optimizations on spatially sparse computations, such as removing element list generation kernels when the topology is unchanged, and demoting data structure activations when the compiler can infer such high-level information. These domain-specific sparse computation optimizations further make way for classical inter-kernel optimizations such as kernel fusion and dead store eliminations. For example, on our MGPCG benchmark (section~\ref{sec:mgpcg}), we find that our optimizations lead to $5\times$ fewer GPU kernel launches, and on our MLS-MPM~\cite{hu2018moving} benchmark, our optimizer is able to automatically fuse the G2P and P2G kernels, leading to an efficient G2P2G kernel~\cite{wang2020massively}.}

We summarize our contributions as follows:
\begin{enumerate}
    \item \revision{A domain-specific data-flow model to analyze imperative spatially sparse computation. The resulted {\em state-flow graphs (SFGs)} serve as a high-level intermediate representation (IR) of imperative, parallel, and spatially sparse programs.}
    \item An asynchronous task execution engine that exposes inter-kernel optimization opportunities and enables parallel compilation;
    \item Most importantly, \revision{{\em an inter-kernel optimizer} for asynchronous spatially sparse computation.} The optimizer can conduct domain-specific optimizations such as list generation removal and sparse data structure activation demotion. Meanwhile, it can also carry out general-purpose inter-kernel optimizations such as dead store elimination and kernel fusion;
    \item A systematic study of the resulted system. Based on the benchmarks, we show our inter-kernel optimizer delivers $1.87\times$ (geometric mean) wall-clock time improvements and $4.02\times$ fewer GPU kernel launches compared to the reference system~\cite{hu2019taichi}. All these can be achieved {\em without} the programmer modifying a single line of kernel code.
\end{enumerate}

We partially reuse the compiler infrastructure of Taichi~\cite{hu2019taichi}. Our source code is attached to the submission, and key files in the compiler implementation are listed in the supplemental document.

\section{Related Work}

\paragraph{Spatially sparse computation} The idea of leveraging spatial sparsity in graphics originates from popular sparse data structures including VDB~\cite{museth2013openvdb, hoetzlein2016gvdb, wu2018fast} and SPGrid~\cite{setaluri2014spgrid, gao2018gpu}. While these data structures have demonstrated effective computation and storage benefits over dense arrays, writing programs that leverage them is not an easy task. Taichi~\cite{hu2019taichi} provides a language abstraction that allows using these data structures as if they are dense, and runtime systems that automatically handle parallel voxel iteration and memory management. These designs benefit the end users, but may end up with more computation. These redundant jobs would need an inter-kernel analysis system to optimize.

Another thread of work on sparse computation is sparse linear algebra languages, such as TACO~\cite{kjolstad2017tensor, chou2018format}, which can effectively generate kernels for Einstein summations on sparse matrices and tensors. Instead of explicitly building the sparse matrices, some simulators use matrix-free computations, which are often the more effective ways for high-performance linear algebra solves in physical simulations (see, for example ~\cite{liu2018narrow}).

\paragraph{Sparse tensors in deep learning frameworks}
\revision{Sparse tensors in ML frameworks such as PyTorch are typically implemented in sparse matrix formats such as the Compressed Sparse Row (CSR) and the ``Coordinate" (COO) formats (e.g., \textcolor{blue}{~\href{https://pytorch.org/docs/stable/sparse.html}{torch.sparse}}). Just like dense tensors in deep learning frameworks, sparse tensors there are immutable. More importantly, sparse tensors in deep learning frameworks have a fixed topology, and the sparse matrix formats lack efficient support for random accesses. These features make operations on them easier to analyze and optimize. In contrast, Taichi supports hierarchical sparse tensors with dynamic topology, which needs a domain-specific dataflow model to optimize.}

\paragraph{Array data-flow analysis} The static-single assignment (SSA) form has been a very popular IR structure. SSA forms are designed for scalar variables, and it cannot directly represent array states, where partial updates may happen. Array SSA forms have been proposed and successfully adopted in parallelization~\cite{knobe1998array} and array privatization ~\cite{maydan1993array}. However, related work in this topic is mostly focused on dense arrays. Our high-level IR system represents not only array partial updates, but also the topology changes in sparse arrays.

\paragraph{Whole-program optimization (WPO)} WPO is also known as Inter-procedural optimization (IPO). For ahead-of-time compilation, IPO typically happens at link time, so sometimes it is also called link-time optimization (LTO). Many existing compilers, such as gcc, MSVC, and clang, already support LTO and WHO (see, e.g., gcc WHOPR~\cite{briggs2007whopr}). While WPO is extensively explored in classical compiling systems, it is still underexploited for spatially sparse computation. The unique computational pattern in sparse computation brings higher complexity and the need for a unified high-level intermediate representation for analysis and optimization.

\paragraph{Computational graph optimization in deep learning frameworks}
A feed-forward deep neural (DNN) network can be naturally represented as directed acyclic graphs (DAG). This leads to a straightforward mapping between DNNs and the computational graph: {\em immutable, dense feature maps} directly map to the graph {\em edges}, and {\em operators} (such as convolutions, max pooling, and element-wise add) maps to the graph {\em nodes}. Consequently, modern deep learning frameworks (TensorFlow~\cite{45381}, PyTorch~\cite{NEURIPS2019_9015}, ONNX~\cite{bai2019}, Theano~\cite{team2016theano}, MLIR~\cite{9370308}) have widely adopted the computational graph to represent the DNN models.  High-level optimizations on the computational graph have been a popular feature in deep learning frameworks. The HLO IR of XLA and PyTorch GLOW~\cite{rotem2018glow} are representative examples. Based on the computational graph, traditional computer optimizations such as operator fusion, dead code elimination (DCE), common subexpression elimination (CSE) can be applied. Tensor Comprehensions~\cite{vasilache2018tensor} and Stripe~\cite{zerrell2019stripe} accelerate deep neural networks using the polyhedral model and graph optimizations such as fusion. We refer the readers to ~\cite{li2020deep} for a good survey.

In deep learning frameworks such as TensorFlow~\cite{45381}, every operation creates a new, immutable buffer (``tensor''), and DNNs are essentially data flow of feature maps. In graphics applications, however, we have to adopt an imperative programming paradigm and support in-place updates, mostly because graphics programmers have been accustomed to using imperative programming (e.g., C++, CUDA, and GLSL) for decades.

Our system is similar to these systems in that a high-level graph-based IR is used, yet the high-level IR must consider its partial updates, sparsity, and ``megakernel'' (i.e., many-in-many-out, hundreds of instructions per kernel) natures of sparse computation code.

\paragraph{GPU code optimization}
Extensive research has become done on code optimization for GPUs. For example, Hong et al. \shortcite{hong2018gpu} optimize SASS via emulation and identifying bottlenecks. Filipovi{\v{c}} et al. \shortcite{filipovivc2015optimizing} optimize CUDA kernels via kernel fusion on operations in the forms of map, reduce, and demonstrated speed up on dense BLAS operations. Bo et al. ~\shortcite{qiao2018automatic} proposed an automatic fusion framework for image processing DSLs. However, an on-the-fly inter-kernel optimization system on GPU imperative programming models that provides maximum flexibility for general-purpose computation is still missing.

\paragraph{Physical Simulation DSLs} Developing high-performance physics solvers is challenging\changed{}{,} and a lot of low-level engineering is needed to exploit the capabilities of modern parallel processors. High-level DSLs models simulations as meshes (Liszt~\cite{Devito:2011:LDS}), sparse linear algebra~\cite{Kjolstad:2016:SLP}, and relational data models~\cite{Bernstein:2016:EDP}.

A lower-level system that is more closely related to our work is the Taichi programming language~\cite{hu2019taichi}. Taichi is a DSL with first-class support for sparse data structures.
In the next section, we briefly cover core Taichi features related to this work.

\section{Taichi background}

Taichi~\cite{hu2019taichi} is a programming language for spatially sparse and differentiable visual computing. As a domain-specific language embedded in Python, Taichi's just-in-time compiler transforms compute-intensive kernels (``Megakernels'', similar to a \lstinline{__global__} GPU kernel in CUDA) into parallel executables. Users can flexibly launch the kernels using Python. 

We refer the readers to ~\cite{hu2020taichi} for an overview of the Taichi programming language. Key Taichi features related to this work are described below.

\subsection{Data-oriented programming} 
\lstinline{field} is a key concept in Taichi that represents data. A \lstinline{field} is essentially a one- to eight-dimensional tensor. Each element of the tensor can be a scalar (e.g., density), a small vector (e.g., velocity), or a small matrix (e.g., stress tensor). Externally, a \lstinline{field} in Taichi is {\em flat}: field elements are always accessed via an \lstinline{x[i, j, k]}-style syntax, regardless of its data layout. Internally, however, data are organized in {\em hierarchical} tree structures, described via structural nodes (SNodes). \revision{The leaf layer stores the actual numerical value, while the intermediate layers act as containers storing the cells of the next layer. Note that there is a duality between a container and a cell: The cell of an intermediate layer becomes the container of its next layer~\footnote{More details are in the ``Data structure organization" section of \textcolor{blue}{~\href{https://taichi.readthedocs.io/en/stable/internal.html}{Taichi internal design documentation}}.}.} See Fig.~\ref{fig:listgen} as an example.

Commonly used SNodes in Taichi are \lstinline{dense}, \lstinline{bitmasked}, \lstinline{pointer}, and \lstinline{dynamic}~\cite{hu2019taichi}. They can easily compose into complex data structures that are dense or sparse.

Note that the field shapes are known at \changed{compile time}{compile-time}, allowing the compiler to easily conduct alias analysis.

\subsection{Spatially sparse programming}

A \lstinline{field} in Taichi can be either dense (similar to a CUDA array) or spatially sparse (such as a VDB~\cite{museth2013vdb} or SPGrid~\cite{setaluri2014spgrid}). The support for spatial sparsity~\cite{hu2019taichi} is a unique feature of Taichi. Most 3D graphics data (especially those stored on the voxel grids) are spatially sparse, and Taichi has first-class support for sparse data structures to leverage this property for acceleration. 

To make sparse data structures as intuitive to use as dense data structures, various designs are made on the syntax, the compiler and the runtime:

\begin{enumerate}
\item{\em Sparse struct-for loops} allow users to iterate over the {\em active voxels} of the sparse data structures conveniently. For example, the following code loops over a 3D sparse field:
\begin{lstlisting}
for i, j, k in x:
    x[i, j, k] += 1
\end{lstlisting}

While iterating over only active parts of a sparse data structure improves performance, implementing such an iteration in parallel is highly non-trivial, since the sparse data structure trees are often highly unbalanced. \revision{Instead of recursively looping on the tree, Taichi handles this by a process called {\em list generation}.}

\revision{\item{\em List generation}: The Taichi runtime maintains a list of active elements for each SNode materialized in the program. During the sparse field traversal, Taichi will re-generate the content of the lists from top to bottom. The list generation procedure at each layer is implemented as a unique GPU kernel. For a given layer, the list generation procedure takes as input the parent SNode list, finds out the active cells in each container in the parent's list (the {\em mask state}), and appends them to the list of the current layer. By induction, when this process is carried out for all layers, we will have all the active voxels of the sparse field.}
Such {\em list generation} (Fig.~\ref{fig:listgen}, left branch) procedure is the key mechanism to achieve load-balanced parallel sparse-for loops\cite{hu2019taichi}.

\begin{figure}[ht]
    \centering
    \includegraphics[width=0.99\linewidth]{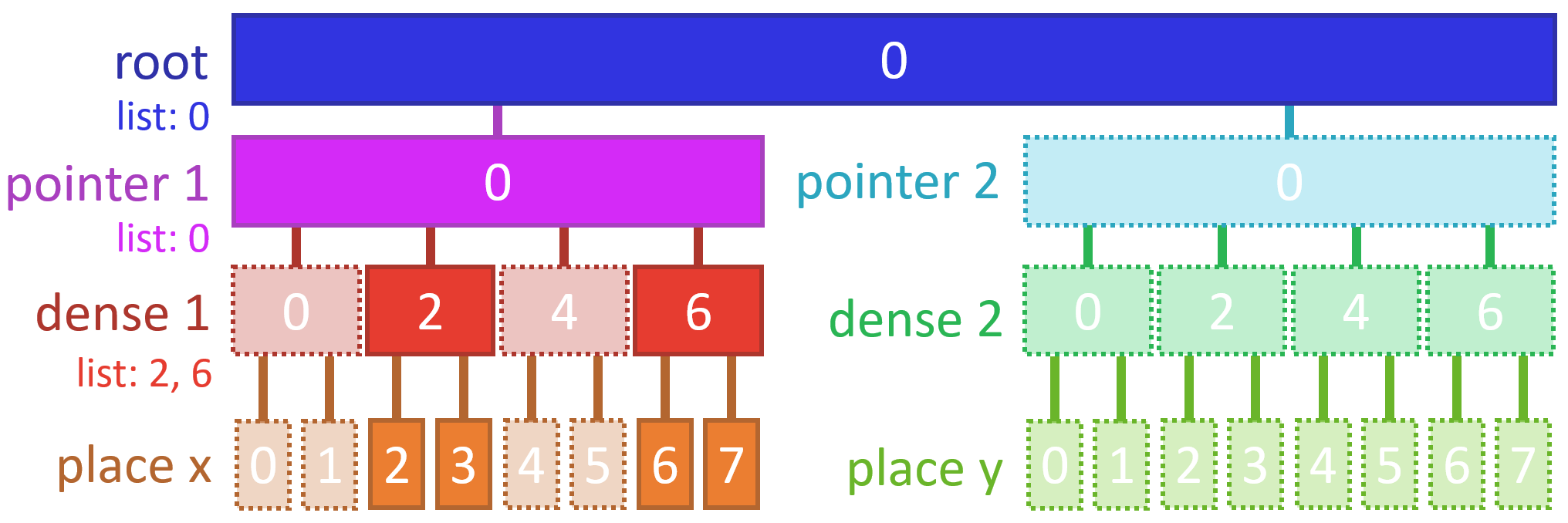}
    \caption{\textbf{Left branch:} The structure of \lstinline{ti.root.pointer(ti.i, 4).dense(ti.i, 2).place(x)} in Taichi, a two-level 1D sparse data structure. The first level, \lstinline{pointer(ti.i, 4)}, is a four-cell pointer array. Each cell of the pointer array can be a null pointer if it is inactive. The second level, \lstinline{dense(ti.i, 2)}, is  dense blocks with two cells each. Highlighted cells are active. Lists of each layer are defined to be collections of active node indices. \textbf{Right branch:} The same structure for \lstinline{field y}, which is completely inactive for now.}
    \label{fig:listgen}
\end{figure}

\revision{The list generation kernel does certain expensive atomic operations such as checking for voxel activation and appending to the list.} In certain cases, the time it takes to generate the lists is comparable to that of the essential parallel iteration. \revision{It is thus critical for Taichi to be able to identify and eliminate the excessive list generation procedures, when the topology of the hierarchy is not changed between the sparse struct-for iterations.}

\item{{\em Activation on write}} ensures sparse data structure nodes are implicitly activated on writing. For example, the following code generates a $2\times 2\times 2$ downsampled sparse field \lstinline{y} from a higher-resolution sparse field \lstinline{x}:
\begin{lstlisting}
for i, j, k in x:
    y[i // 2, j // 2, k // 2] += x[i, j, k]
\end{lstlisting}
Note that the corresponding voxels of \lstinline{y} may not be active before this for loop. Taichi will automatically activate \lstinline{y[i // 2, j // 2, k // 2]} and zero-fill the voxels. See Figure~\ref{fig:activate-on-write} for an example.

\begin{figure}[h]
    \centering
    \includegraphics[width=0.9\linewidth]{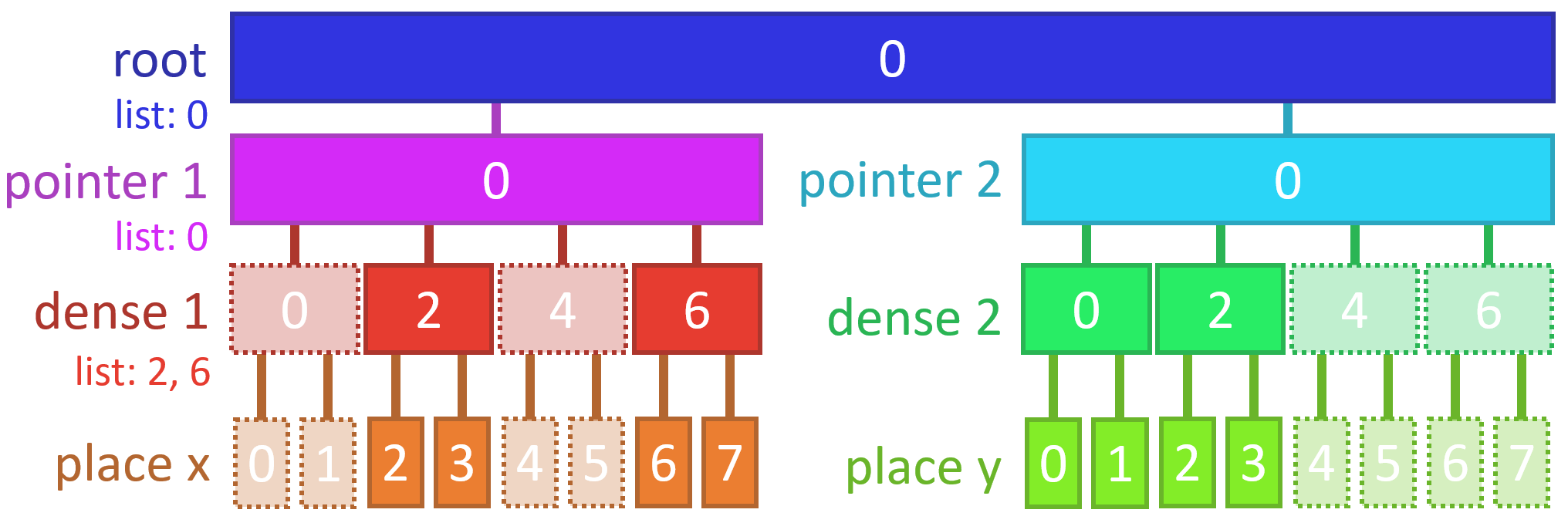}
    \caption{Execution result of simple program \lstinline{for i in x: y[i // 2] += 1}. \lstinline{y[1]} and \lstinline{y[3]} are activated on write. Because dense nodes cannot be only partially active, \lstinline{y[0]} and \lstinline{y[2]} are also activated.}
    \label{fig:activate-on-write}
\end{figure}

\end{enumerate}

\if(0)
\item{\em Automatic parallelization}. Taichi {\em kernels} are decomposed into {\em tasks} that are serial or parallel. For example, in the following code, the for loops are automatically parallelized:
\begin{lstlisting}
@ti.kernel
def reduce():
    s = 0 # Serial
    for i in range(128):
        s += x[i] # Parallel range-for
    for i in y:
        s += y[i] # Parallel struct-for
    print(s) # Serial
\end{lstlisting}
\fi

\section{A State-flow formulation of Imperative Sparse Computation}
\label{sec:sfg}

In imperative programming, {\em Program = State + Compute}.
\revision{Compared to traditional computation on dense arrays, in spatially sparse computation, ``state'' refers to not only the numerical values of \lstinline{field}s, but also the auxiliary data structures to support sparsity. Similarly. ``compute'' also means more than GPU kernels that operate on voxel data, because iterating over sparse data structures in parallel implicitly leads to auxiliary computations such as list generation and node activation. These additional complexity create more challenges in modeling and analyzing spatially sparse programs.}

For now, let us assume we know the whole execution history of a Taichi program. \revision{Drawn inspiration from traditional data-flow analysis,} we reformulate the imperative computation scheme of Taichi into a collection of {\em states} and {\em tasks}.
To systematically optimize imperative GPU programs, especially those with spatial sparsity support, we formulate a Taichi program as a \textbf{state-flow graph (SFG)}, \revision{a domain-specific data-flow graph on spatially sparse computation. As a data-flow graph, SFG is a directed acyclic graph (DAG) with nodes being {\em tasks} and edges being {\em states}.} This results in a state flow formulation and a high-level intermediate representation (IR). Scalar data-flow analysis is well studied in optimizing compilers (see, for example, ~\cite{khedker2017data}), and SFG \revision{is an extension of} data-flow analysis to handle auxiliary states such as lists in spatially sparse computation.

\begin{figure}
    \centering
    \includegraphics[width=1.0\linewidth]{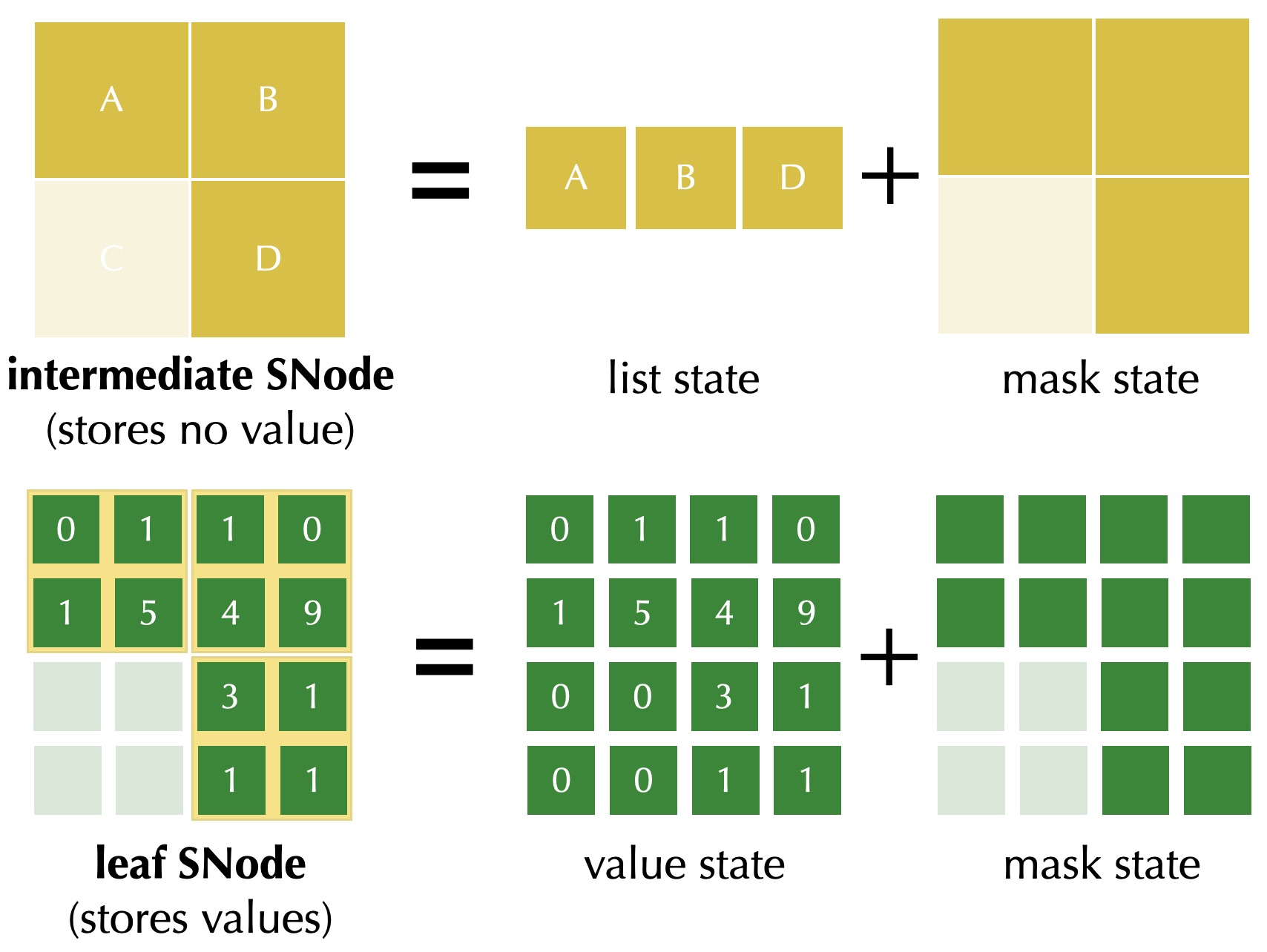}
    \caption{State decomposition of a two-level sparse array, containing a sparse intermediate layer and a dense leaf layer. Note that the value state covers all pixels, even if the pixel is inactive. In other words, whenever an access reads a pixel from the sparse array, the mask state will first be queried. If the mask state says the pixel is inactive, $0$ will be returned. Otherwise\changed{}{,} the system queries the value state and returns the corresponding value. Here we omit allocator states for simplicity.}
    \label{fig:states}
\end{figure}

\paragraph{States} States split the holistic description of a Taichi program into a suitable granularity for analysis and optimization. For SNodes that are spatially sparse, we must decompose the holistic descriptions of their {\em data, topology, and auxiliary structures} into the following kinds of states:

\begin{itemize}
    \item{\em A value state} simply represents the collection of numerical values stored in the field. Note that in the data structure trees of Taichi, only the leaf nodes (i.e., \lstinline{place} SNodes) store numerical values. \revision{Value states are the most basic states. They have the same meaning as those in data-flow analysis, and are useful in almost all GPU programming systems.} It is worth noting that in sparse data structures, every voxel has a numerical value, even if the voxel is inactive - in that case, the inactive voxel has an ambient value $0$.

    \item{\em A mask state} of a SNode records the activation information of all its cells (Fig.~\ref{fig:states}, right). Mask states must be handled as first-class primitives in our SFG system for domain-specific optimizations. \revision{Mask by itself is not a piece of data materialized in the memory, but a unified abstraction whose state could be inferred via different ways for each kind of the sparse SNode. For example, a \lstinline{bitmask} SNode maintains a list of integers as the bitmask metadata, one bit for each cell. Therefore, the active mask contains those cells whose corresponding bits are activated. Another example is the \lstinline{pointer} SNode. The elements of such a SNode are just pointers, hence non-null slots comprise the active mask state.} \SOUT{Masks are scattered in various forms in the data structure. They are either indicated by a bit in a bitmask SNode, or a non-null pointer for the pointer SNode. Mask is a unified concept for different data structures, sparse or dense.}

    \item{\em A list state} of an SNode represents the data structure nodes maintained by the runtime system. Recall that Taichi needs to generate/consume data structure node lists for load-balancing parallel iterations over sparse data structure nodes. \revision{List generation tasks take the mask state of the current SNode and the list state of the parent SNode to generate the list of the current SNode.} Lists are consumed by (parallel) struct-fors. See~\cite{hu2019taichi} for more details on load balancing and parallel fors on unbalanced trees.
    
    \item{\em An allocator state} represents the state of Taichi's memory allocator. For computation that allocates/deallocates sparse data structure nodes, the allocator states are marked as modified.
\end{itemize}

\revision{A state is tagged with a version number. Every time a state is modified within a task, its version number is incremented, with the underlying data buffer of the SNode mutated in-place. Doing so allows Taichi to track the latest writer (owner) of a given state, thereby enables many opportunities for optimization (more on this in section \ref{sec:optimization}).}

The relationship between value, mask, and list states is depicted in Fig. ~\ref{fig:states}.

\paragraph{Tasks} A Taichi kernel may be decomposed into multiple parallel tasks (GPU kernels). Without loss of generality, we assume that a Taichi kernel corresponds to a single task and generates a single GPU kernel\footnote{We use the term ``task'' and ``kernel'' interchangeably for a serial/parallel execution job on GPUs.}. Each Taichi task has input edges (input states), output edges (modified states). It also maintains its metadata, such as loop ranges.
These edges and metadata will be used for inter-kernel optimization.

\subsection{State-flow chains}

\begin{figure}[h]
    \centering
    \includegraphics[width=1.0\linewidth]{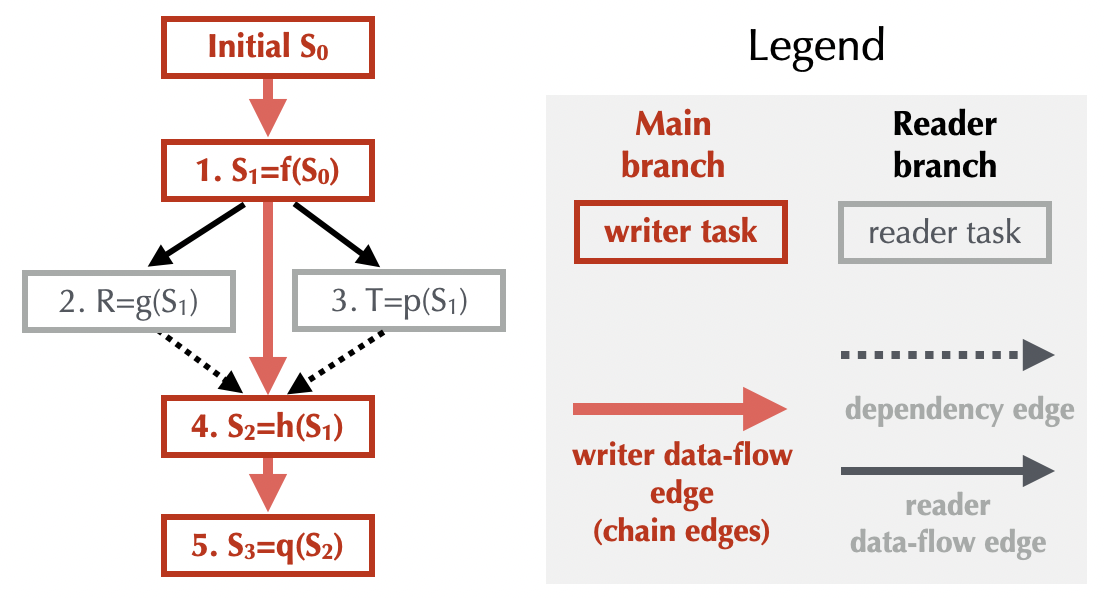}
    \caption{A state-flow chain of value state $S$. The edges in the state-flow chain depict the task dependency relationships. Note that each state-flow chain always has a main branch (write-after-write, red in the figure) and a few reader branches (read-after-write \& write-after-read). On the main branch, each node (task) creates a new version of the state. We classify write-after-write and read-after-write as data-flow \changed{edge}{edges}, since there are data produced and consumed. Write-after-read edges are classified as dependency edges.}
    \label{fig:chain}
\end{figure}

Now let us focus on a single state. For example, we use value state $S$ (Fig.~\ref{fig:chain}), which is manipulated by kernels (tasks) $f, g, p, h, q$. Note that $f, h$ and $q$ read and write the value state $S$, yet $g$ and $p$ only reads the value of $S$. Clearly, only the latest writer holds the most up to date version of a state, while readers only fetch the latest state \revision{without resulting in a new version}. If we only consider the writers, we get a chain structure for each state, with a few branches for readers. Fig.~\ref{fig:chain} provides a concrete example.

It is worth pointing out that the $\phi$ node in the SSA form is not needed in SFC. In the execution model of Taichi, the kernel-level control flow is directly evaluated inside the host language (Python), rather than being part of the DAG.

For a single state\changed{}{,} we can easily build a chain (which is also a DAG). We call the chain structure a ``state-flow chain'' (SFC).

\subsection{State-flow graphs}

\begin{figure}[h]
    \centering
    \includegraphics[width=1.0\linewidth]{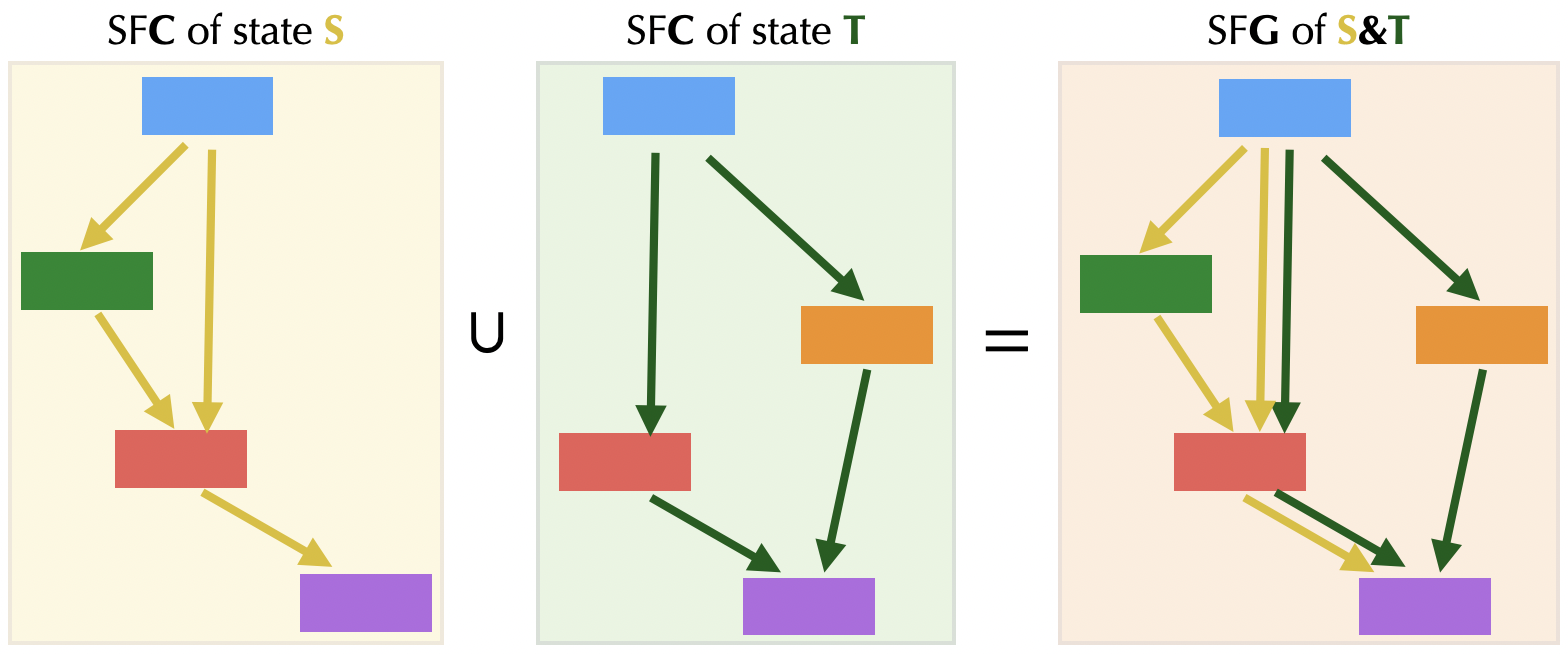}
    \caption{A state-flow graph, by definition, is a union of state-flow chains of all the states used in a program. Note that each edge represents a state and a node represents a task. Two tasks may be connected by more than two edges, each edge representing a state.}
    \label{fig:graph}
\end{figure}

A Taichi program can easily have hundreds of states. Here we introduce state-flow graphs (SFGs), which are essentially state-flow chains sticking together, or unioning their nodes and edges (Fig.~\ref{fig:graph}). SFGs completely describe the relationship between tasks in Taichi. Since unions of DAGs following the same topological order are still DAGs, SFGs are DAGs too. 

The SFG serves as the IR for inter-kernel optimizations. The SFG formulation allows us to use well-established graph theory languages for compiler optimization. For example, our task fusion optimization uses reachability analysis in graphs (section~\ref{sec:fusion}). 

Whenever a task is inserted into the execution queue, we dynamically create an SFG node and create the corresponding dependency edges. SFGs have two useful properties:

\begin{enumerate}
\item{\textbf{Order independency}.} Any topologically ordered task sequence leads to the same program behavior.
\item{\textbf{Reconstruction invariance}, corollary of ``order independency''.} Any topologically ordered task sequence of G constructs the same graph G.
\end{enumerate}

``Reconstruction invariance'' is particularly useful when manipulating the graph nodes. For example, to remove a node from SFG, simply topologically sort the SFG nodes, remove the node from the sorted list, and rebuild the SFG. This frees us from worrying about how to handle edges that are connected to the removed node, or to update the latest set of owners of the affected states in the system.

\section{Lazy and asynchronous kernel launches}

In existing parallel programming languages such as CUDA, kernels are ahead of time (AOT) compiled and launched immediately once called on \changed{}{the} host\footnote{Existing GPU programming systems such as CUDA and OpenGL already provide some asynchrony between the CPU host and GPU devices, but we need more asynchrony for inter-kernel optimizations, as described later in \changed{the}{this} section.}. 
However, we need two more execution mechanisms to make inter-kernel optimizations work: just-in-time (JIT) compilation \changed{compilation}{} and (kernel-level) lazy evaluation.

\paragraph{JIT compilation} The issue with AOT compilation is that, at launch time, optimizers only have access to low-level assembly code (e.g., PTX or SASS)\changed{}{,} which is too fine-grained and fragmented for further optimizations. Fortunately, Taichi not only provides a JIT system, but also allows to lower the IR halfway to a level that is very suitable for inter-kernel optimizations (see Fig.~\ref{fig:teaser}, bottom left, where inter-kernel optimizations happen at the middle-end IR).

\paragraph{Asynchronous launching} In CUDA\changed{}{,} GPU kernel execution is asynchronous, but GPU kernel launches are still eager. The eager launching mechanism prevents cross-kernel optimization \changed{to happen}{from happening}, since the system only {\em sees one kernel at a time}. Therefore, \changed{in order to}{to} make the SFG practically useful, we need to hold the SFG nodes from executing before inter-kernel optimizations.

We developed an {\em asynchronous} execution engine for GPU programs. The existing Taichi system eagerly launches the kernels, but we can modify the system, making it {\em asynchronous}, and maintain a list of kernels to compile and run {\em lazily}. This opens up opportunities for inter-kernel optimizations detailed in the following section.

\paragraph{By-product: parallel compilation} A drawback of JIT compilation is its compilation time. Note that ahead of time compilation does not have this issue. In fact, as Taichi becomes more widely adopted, the compiler needs to deal with programs with increasing instructions and optimization passes, in extreme cases compilation can \changed{sometimes} take up to $70\%$ of program end-to-end run time. In the previous eager execution scheme, a serial thread is used to compile and launch these kernels. In contrast, since the asynchronous execution engine sees multiple kernels at a time, parallel compilation can be done easily, which can significantly reduce wall-clock time spent on the compilation. The effectiveness of parallel compilation is evaluated in section~\ref{sec:timeline}.

\section{Optimize across kernel boundaries}
\label{sec:optimization}

With the state-flow graph IR that describes the task relationships, and the asynchronous execution engine that saves the tasks from being executed too early, we can finally conduct analysis and optimizations on the state-flow graph. In this section, we discuss four effective inter-kernel optimizations on spatially sparse computation programs.

\subsection{A minimal example}

\revision{
Here we show a trivial optimization example of two Taichi kernels.
}

\begin{figure*}[t]
    \centering
    \begin{tabular}[t]{cc}
        \begin{subfigure}{0.63\linewidth}
            \centering
            \includegraphics[width=\linewidth]{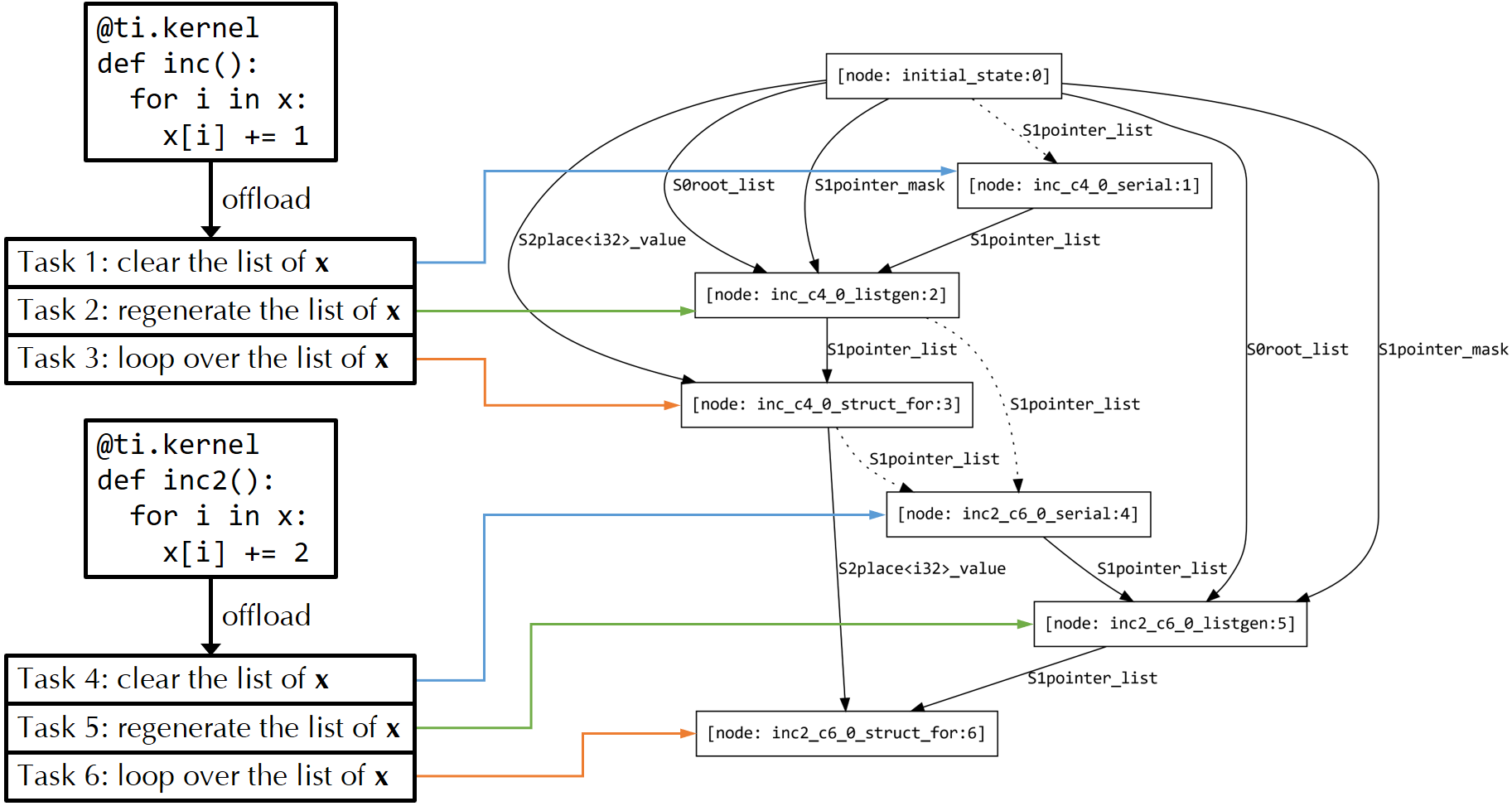}
            \caption{Generated tasks from two Taichi kernels and the corresponding SFG.
		\revision{\textbf{x} is a sparse \lstinline{field}, and the structure is \lstinline{ti.root.pointer(ti.i, n).place(x)}. Each task corresponds to a node in the SFG. The labels of the nodes in the SFG shows the task type of the node (except for the initial state, which does not correspond to a task), where ``serial'' denotes serial tasks including clear list tasks, ``listgen'' denotes list generation tasks, and ``struct\_for'' denotes struct-for tasks.
		In the labels of the edges in the SFG, ``S0'' corresponds to the root SNode, ``S1'' corresponds to the pointer SNode, and ``S2'' corresponds to the place SNode of \textbf{x}.}}
            \label{inc-initial}
        \end{subfigure}
        &
        \hspace{-3mm}
        \begin{tabular}{c}
        	\begin{subfigure}[t]{0.36\textwidth}
        	\includegraphics[width=\linewidth]{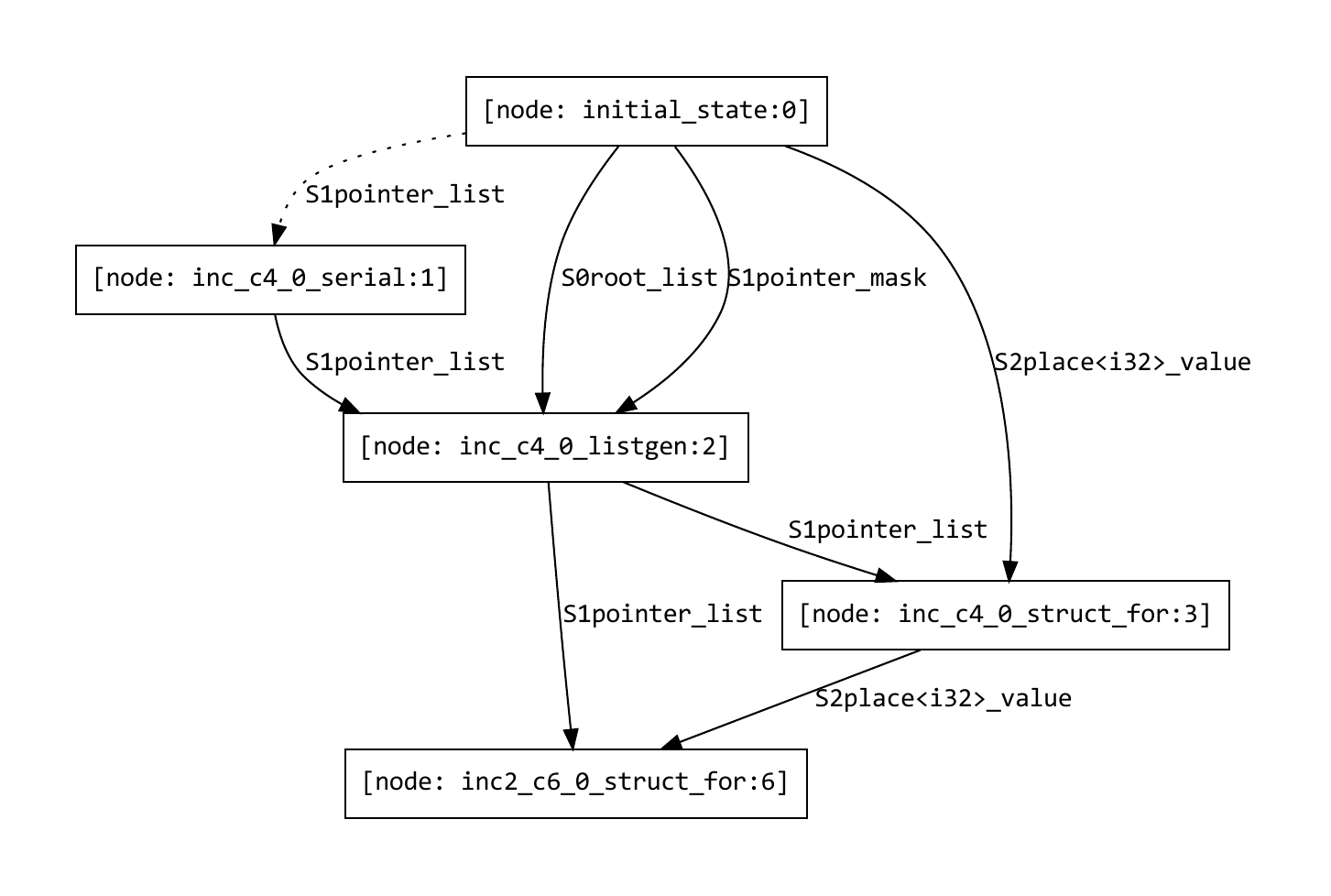}
        	\vspace{-5mm}
        	\caption{\revision{The SFG after list generation removal. The node ``inc2\_c6\_0\_listgen:5'' and the serial (clear list) node preceding it are removed.}}
        	\label{inc-sfg-listgen}
        	\end{subfigure}\\
        	\begin{subfigure}[b]{0.36\textwidth}
        	\includegraphics[width=\linewidth]{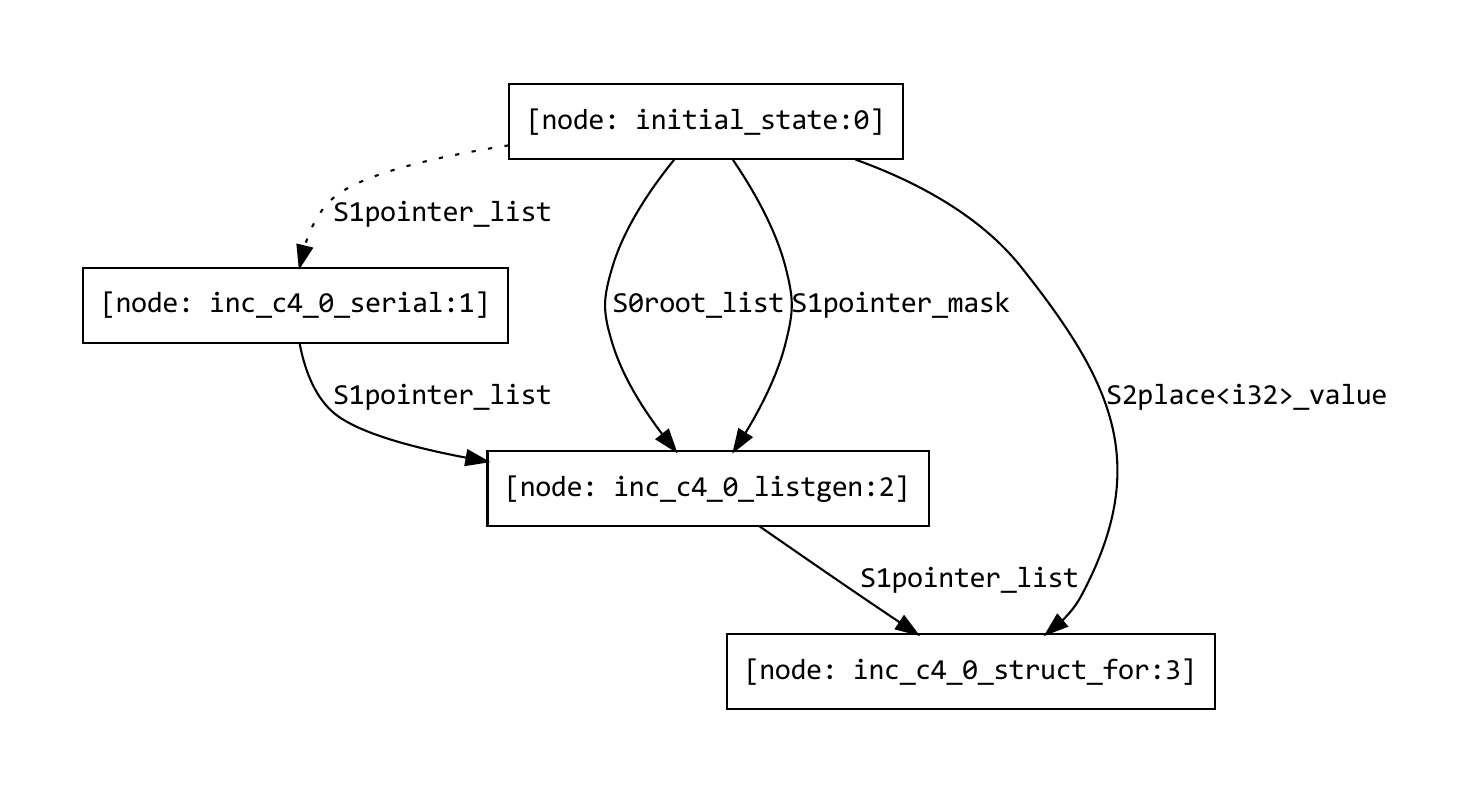}
        	\vspace{-5mm}
        	\caption{The SFG after task fusion. \revision{The node ``inc2\_c6\_0\_struct\_for:6'' is fused into ``inc\_c4\_0\_struct\_for:3''.}}
        	\label{inc-sfg-fuse}
        	\end{subfigure}
    	\end{tabular}
	\end{tabular}
	\caption{State-flow graph optimizations. (a) demonstrates the correspondence between Taichi kernels and the SFG. }
	\label{sfg_opt}
\end{figure*}

\begin{figure}[ht]
	\centering
	\includegraphics[width=0.99\linewidth]{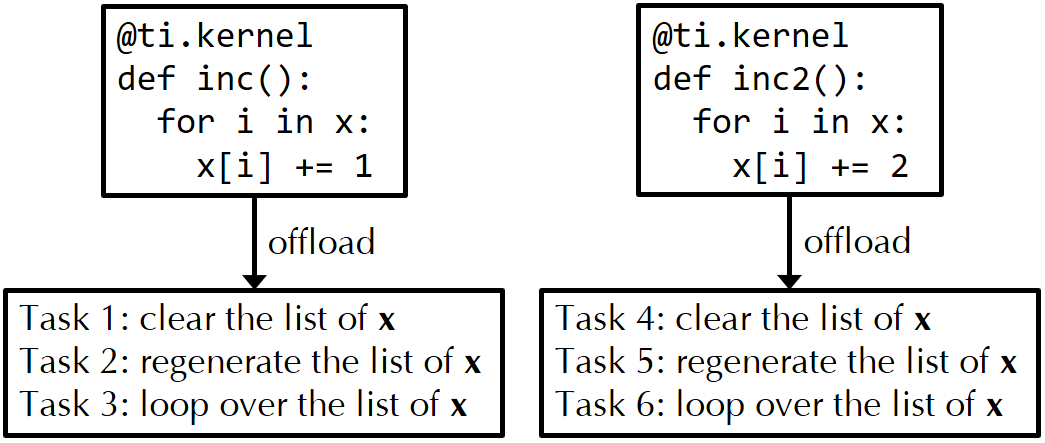}
	\caption{The generated tasks of two kernels without kernel fusion.
		\textbf{x} is a sparse \lstinline{field}.}
	\label{without_fusion}
\end{figure}

\begin{figure}[ht]
	\centering
	\includegraphics[width=0.99\linewidth]{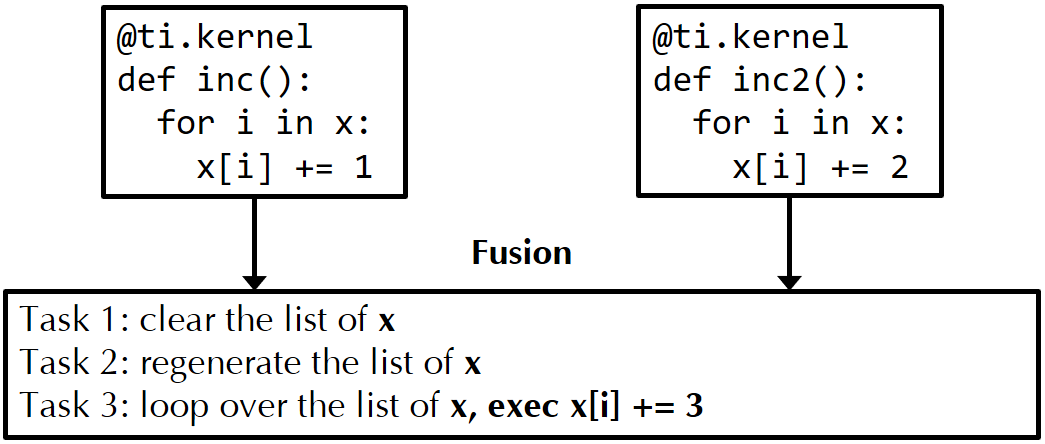}
	\caption{The generated tasks of two kernels with kernel fusion.}
	\label{with_fusion}
\end{figure}

\revision{
As shown in Fig.~\ref{without_fusion}, since \textbf{x} is a sparse data structure, Taichi needs to generate an active list of \textbf{x} to know which elements of \textbf{x} need to be looped over. So there are 3 tasks per kernel in the original Taichi system. 
Note that the kernels are lightweight here, the running time of list generation tasks is comparable to the essential computation time. If the kernel on the right succeeds the kernel on the left of Fig.~\ref{without_fusion}, in our system,
we can perform some analysis to know that the mask state of \textbf{x} is not changed, fuse the two kernels into one, and finally get Fig.~\ref{with_fusion} after optimizations.
}
In this case, we reduce the number of generated tasks from 6 to 3. Kernel fusion is not new, but fusing kernels that \changed{operates}{operate} on {\em sparse} data structures is a unique challenge in Taichi, since the iteration over active elements implicitly depends on the mask of the sparse data structures.

Even if the bodies of both kernels cannot be optimized in the same way as this example, we can still remove some list generation tasks and reduce running time. This can be a significant improvement for small kernels where the list generation time is comparable to the real computation time.

Common GPGPU patterns and Taichi's sparse computation model motivates us to apply the following domain-specific and general-purpose compiler optimizations:
\begin{itemize}
    \item List generation removal
    \item Activation demotion
    \item Task fusion
    \item Dead store elimination
\end{itemize}

The remainder of this section details these optimizations.

\subsection{List generation removal}
This is the easiest whole program optimization, yet it leads to significantly higher performance for sparse computations in certain cases. \revision{A list generation task is idempotent in the sense that, if its input parent list state and the mask state is the same, it will always produce the same list state. Because of this property, if no modifying task is launched between two struct-for loops over the same SNode, the state versions in the SFG will stay the same. Thus the list generation tasks associated with the later loop can be safely eliminated.}

List generation removal not only saves unnecessary execution time on generating the sparse element lists, but also opens up opportunities for other optimizations. For example, if two struct-for tasks are using the same list after list generation removal, a {\em task fusion} may fuse the tasks.

\subsection{Activation demotion}

Recall that Taichi has an activation-on-write mechanism. It is often the case that the sparse element was already activated before the task execution, so the element activeness was checked \changed{by not re-activated}{to avoid unnecessary activation}. This extra check not only creates diverging instruction flow on CPU/GPUs that harms performance, but also creates a modification to the corresponding mask state, creating obstacles for list generation removal. Therefore, we should try to demote activating accesses to non-activating accesses. 

Fortunately, many activations can be demoted, by analyzing the task contexts. If two struct-for tasks are identical, the loop lists are the same, and the activation statement in the second task depends only on the loop indices, then the activation in the second task can be removed.

This optimization is remarkably effective for repeated access patterns such as \lstinline{[i // 2]}. For example, in the restriction (downsample) operator of multigrid solvers, it is common to have the following pattern (Fig.~\ref{downsample}):
\begin{lstlisting}
for i, j in x:
    y[i // 2, j // 2] += x[i, j] * 0.25
\end{lstlisting}

\begin{figure}[ht]
    \centering
    \includegraphics[width=0.6\linewidth]{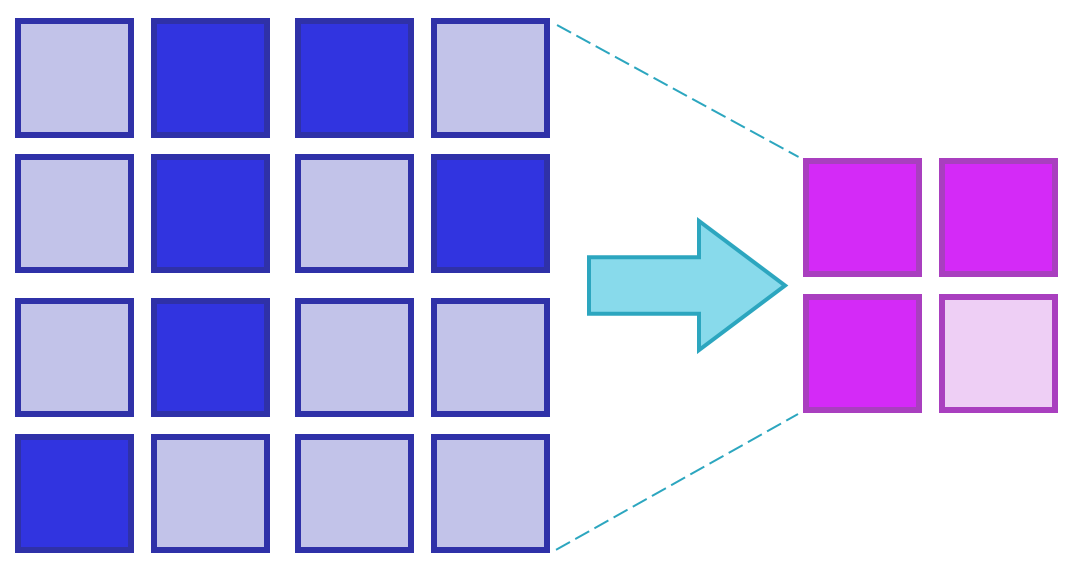}
    \caption{The activation pattern of \lstinline{for i, j in x: y[i // 2, j // 2] += x[i, j] * 0.25}. \lstinline{x} is the grid on the left, and \lstinline{y} is the grid on the right.}
    \label{downsample}
\end{figure}

Our activation elimination optimizer can successfully infer that if the mask of $x$ has not been changed, then the mask of $y$ will not change either. This \changed{avoid}{avoids} false-positive mask state modifications, and can further bring down the list generation kernel tasks by $6.7\times$ in the MGPCG example.

\subsection{Task fusion}
\label{sec:fusion}
\revision{Task fusion is an effective optimization to improve locality and reduce the number of kernels launches.
Without list generation removal, we cannot fuse the two struct-for tasks in Fig.~\ref{inc-initial} because they take as input two different lists (different versions of the list state of the pointer SNode). Fortunately, after removing the list generation task in Fig.~\ref{inc-sfg-listgen}, we can fuse the tasks, resulting in Fig.~\ref{inc-sfg-fuse}. The pattern of loops in Fig.~\ref{sfg_opt} is very common in spatially sparse programming, and we need list generation removal to open up space for task fusion.}
\revision{We use the SFG to find pairs of tasks to fuse, and this can be time-consuming when there are too many tasks. Please see the supplemental document for details about the fusing criteria and the algorithm for efficiently finding all fusible pairs of tasks.}

\subsection{Dead store elimination}
We can also perform some general-purpose inter-kernel data-flow analysis. For example, \lstinline{ti.clear_all_gradients()} may excessively zero-fill unrelated gradient fields, which can be eliminated with data-flow analysis.

For convenience\changed{}{,} a user may frequently zero-fill fields in Taichi to ensure data are correctly re-initialized. This is a typical source of dead stores. For such cases where a field is completely overwritten, our optimizer can eliminate the previous dead store:

\begin{lstlisting}
@ti.kernel
def clear():
    for i in x:
        x[i] = 0
        y[i] = 0
        
@ti.kernel
def inc_x():
    for i in x:
        x[i] += 1
       
    for i in x:
        print(x[i])

clear()
# After DSE, y[i] = 0 in this kernel is eliminated  

inc_x()
clear()
\end{lstlisting}

\section{Implementation details}
The inter-kernel optimizations are relatively simple to implement, but extra attention was paid to the infrastructure to support these optimizations. In this section, we briefly cover implementation details that we empirically found to directly impact performance.

\subsection{Asynchronous Execution Engine}

We implement an asynchronous execution engine that performs SFG optimizations and parallel compilation. 

All tasks invoked from the Python side are initially accumulated inside a queue, until either an implicit synchronization event (e.g., data transfer between the device and the host), or an explicit call to flush (explained in the next paragraph) happens. Upon such events, the tasks are popped off from the queue to construct an SFG instance. This SFG instance goes through the various kinds of optimizations mentioned in Section \ref{sec:optimization}. Once the tasks in the graph are finalized, they are sent to the JIT compilation workers running in parallel, then the backend device.

\paragraph{Flushing} While the more tasks are deferred, the more information is retained for the SFG to optimize, it is usually undesirable to solely depend on the implicit synchronization events to flush the task queue, since this can easily cause starvation on either CPU or GPU side. To provide experienced users with more control over the asynchronous execution engine, we provide a simple API, \lstinline{ti.async_flush()}, to flush the tasks to the SFG optimizer and then the GPU device. This API is non-blocking, which allows for overlapped execution between CPU and GPU. For \changed{the}{} most of the usage cases, our system is configured to periodically flush the tasks {\em automatically}. While this simple strategy could lead to sub-optimal executions, in practice\changed{}{,} we have found this to yield sufficient performance. Note that setting the flushing period to $1$ effectively turns off the asynchronous execution. 

\paragraph{Partial SFG Garbage Collection} To further mitigate the loss of information potentially caused by flushing or synchronization, each time an optimized SFG instance is sent for execution, Taichi does a partial garbage collection by preserving those nodes that are the latest owners of the states. As new tasks get launched, these nodes will become the roots in the new SFG instance. This enables the system to capture the information it needs for certain \changed{type}{types} of optimizations. For example, assuming one preserved SFG node is the latest owner of an SNode's mask state, and a new sparse struct-for loop task reading that SNode is launched. If the mask state has not been modified in between, the SFG optimizer \changed{is able to}{can} infer that it is safe to remove the list generation tasks preceding the struct-for task.

\subsection{IR handle and IR bank for caching compilation}
\label{sec:irbank}
Since a kernel can be launched many times with the same IR, we store all IRs into an IR bank to avoid repeated passes on the IR and to improve the asynchronous compilation performance. We use IR handles to access IRs in the bank. An IR handle consists of a pointer to the IR and the hash of the IR. We assign an IR handle to each task, and whenever we are going to do any modification to the IR, we check if we have already done it in the IR bank, where we cache the result of IR optimization passes such as fusion, activation elimination, and dead store elimination. If the result is not cached, we copy the IR on write to avoid corrupting the IR in the bank, do the modification, store the modified IR into the bank, and then cache the mapping from the IR handle before modification to the IR handle after modification into the bank. We also cache some data that do not need to modify the IR into the bank, such as the \changed{task meta of the IR}{task's metadata}.

\subsection{Intra-kernel data-flow optimizations}
\label{sec:intra}
To achieve better performance after task fusion, we need an optimization pass on the task after fusion. As Taichi IRs are inherently hierarchical, we build a data-flow graph for data-flow analysis, to perform inter-kernel optimizations, including store-to-load forwarding, dead store elimination, and identical store/load elimination. For example, in Figure \ref{with_fusion}, on CPU we demote atomic addition operations into loads, adds and stores, and with store-to-load forwarding, we can replace the load of the second atomic addition (\textbf{x[i] += 2}) with the addition result of the first atomic addition (\textbf{x[i] += 1}), and get the final result as if the input was \textbf{x[i] += 3} with other optimizations.

More details on intra-kernel data-flow optimizations can be found in the supplemental document.

\section{Evaluation}

In this section\changed{}{,} we systematically evaluate our system on microbenchmarks and large-scale end-to-end test cases.

\paragraph{Metrics} 
On each test case, we evaluate the performance with five metrics:
\begin{enumerate}
\item Wall-clock time (inter-kernel optimizer time included);
\item Backend (GPU or CPU thread pools) execution time;
\item Number of tasks launched;
\item Number of instructions emitted to the code generator;
\item Number of tasks compiled.
\end{enumerate}

Each case is executed multiple times on GPU (CUDA) and CPU (x64), with a synchronization after each run in asynchronous mode, and the total metrics over all runs are recorded.

\paragraph{Benchmark cases} We constructed 10 simple yet indicative microbenchmarks (tens of lines of code each) to unit-test specific inter-kernel optimizations. Four complex test cases (hundreds of lines of code each) test the behavior of our optimizer on real-world programs, including computational physics tasks on regular sparse grids (section ~\ref{sec:advection}), hybrid Lagrangian-Eulerian schemes (particles and grids, section ~\ref{sec:mlsmpm}), multi-resolution sparse grids (section ~\ref{sec:mgpcg}), and triangular meshes (section~\ref{sec:autodiff}).

\begin{table*}[h]
\caption{\label{table:benchmark} Benchmarks against the original Taichi system~\cite{hu2019taichi} without inter-kernel optimizations. The baseline system and applications are tuned against state-of-the-art manually engineered CPU and GPU implementations, as detailed in ~\cite{hu2019taichi}. Benchmarks are done on a system with a quad-core Intel Core i7-6700K CPU with 32 GB of memory, and a GTX 1080 Ti GPU with 12 GB of GRAM. The geometric mean all benchmarks:  the wall-clock speed up is $1.87\times$ (CUDA) / $1.33\times$ (x64) and the reduction of task launched is $4.02\times$. Commands to reproduce all the numbers are included in sections detailing each experiment.}

\begin{tabular}{|c|c|c|r|r|r|r|r|r|r|r|r|r|}
\toprule
\multicolumn{2}{|c|}{\multirow{2}{*}{\textbf{Cases}}} & \multirow{2}{*}{\textbf{Backend}} & \multicolumn{2}{c|}{\textbf{Wall-clock time (s)}}     & \multicolumn{2}{c|}{\textbf{Backend time (s)}}        & \multicolumn{2}{c|}{\textbf{Tasks launched}}          & \multicolumn{2}{c|}{\textbf{Instructions emitted}}    & \multicolumn{2}{c|}{\textbf{Tasks compiled}}          \\ \cline{4-13} 
\multicolumn{2}{|c|}{}                                &                                   & \multicolumn{1}{c|}{Ref.} & \multicolumn{1}{c|}{Ours} & \multicolumn{1}{c|}{Ref.} & \multicolumn{1}{c|}{Ours} & \multicolumn{1}{c|}{Ref.} & \multicolumn{1}{c|}{Ours} & \multicolumn{1}{c|}{\quad Ref. \quad} & \multicolumn{1}{c|}{Ours} & \multicolumn{1}{c|}{\quad Ref. \quad} & \multicolumn{1}{c|}{Ours} \\ \midrule \hline
\multicolumn{2}{|c|}{\multirow{2}{*}{MacCormack}}     & CUDA                              & 8.497                & 2.907                 & 8.479                   & 2.874               & 4726                 & 319                & 16880             & 6277                 & 96                    & 15                    \\ \cline{3-13} 
\multicolumn{2}{|c|}{}                                & x64                               & \multicolumn{1}{c|}{206.115}    & 73.520        & 206.065               & 73.468               & 4726              & 319                  & 16880             & 6277             & 96                & 15                \\ \hline \hline
\multirow{4}{*}{MGPCG}      & \multirow{2}{*}{2D}     & CUDA                              & 9.185                     & 3.690                     & 7.799                     & 2.101                     & 1057560                   & 224568                    & 3387                      & 3816                      & 204                       & 105                       \\ \cline{3-13} 
                            &                         & x64                               & 23.640                    & 20.468                    & 20.970                    & 19.841                    & 1057560                   & 224570                    & 2961                      & 3331                      & 204                       & 106                       \\ \cline{2-13} 
                            & \multirow{2}{*}{3D}     & CUDA                              & 9.352                     & 6.500                     & 8.960                     & 6.244                     & 304392                    & 63869                     & 3652                      & 4450                      & 146                       & 82                        \\ \cline{3-13} 
                            &                         & x64                               & 172.599                   & 161.927                   & 171.135                   & 161.607                   & 304392                    & 63869                     & 2728                      & 3200                      & 145                       & 77                        \\ \hline\hline
\multicolumn{2}{|c|}{\multirow{2}{*}{MLS-MPM}}        & CUDA                              & 14.059                    & 10.633                    & 13.987                    & 10.584                    & 18806                     & 9201                      & 8900                      & 20008                     & 122                       & 91                        \\ \cline{3-13} 
\multicolumn{2}{|c|}{}                                & x64                               & 292.615                   & 280.415                   & 292.307                   & 280.376                   & 18806                     & 9201                      & 9132                      & 20492                     & 122                       & 91                        \\ \hline\hline
\multicolumn{2}{|c|}{\multirow{2}{*}{AutoDiff}}       & CUDA                              & 2.256                     & 1.377                     & 2.124                     & 1.181                     & 65674                     & 42454                     & 1346                      & 2184                      & 22                        & 32                        \\ \cline{3-13} 
\multicolumn{2}{|c|}{}                                & x64                               & 60.947                    & 53.278                    & 59.308                    & 53.159                    & 65674                     & 42454                     & 1353                      & 2174                      & 22                        & 30                        \\ \hline
\bottomrule
\end{tabular}
\end{table*}

\subsection{Microbenchmarks} 
We constructed 10 microbenchmark cases to unit-test the system. The results are promising: without code modification, the new system leads to $3.73\times$ fewer kernel launches on GPUs and $2.5\times$ speed up on our benchmarks. More details on the microbenchmarks are discussed in the supplemental document.

\begin{table}[ht]
\caption{Geometric mean results of the 10 microbenchmark cases. Numbers are ratios between the reference system~\cite{hu2019taichi} and ours with inter-kernel optimizations.}
\begin{tabular}{lrr}
\toprule
\textbf{Metric}   &  \textbf{CUDA} & \textbf{x64} \\ \midrule
\midrule
Wall-clock time    &   $2.30\times$ & $2.14\times$ \\\midrule
Backend time        &   $2.56\times$ & $2.32\times$  \\\midrule
Tasks launched          &   $3.73\times$ & $3.69\times$ \\\midrule 
Instructions emitted     &   $0.97\times$ & $0.97\times$ \\\midrule
Tasks compiled          &   $1.15\times$ & $1.15\times$\\ \midrule
\bottomrule
\end{tabular}
\label{table:micro}
\end{table}

\if(0)
('wall_clk_t', 'cuda', 'sync')
  geometric mean 0.020867636705843143
('wall_clk_t', 'cuda', 'async')
  geometric mean 0.009059276018712231
('wall_clk_t', 'x64', 'sync')
  geometric mean 0.010373569345101204
('wall_clk_t', 'x64', 'async')
  geometric mean 0.004847401045842292
('exec_t', 'cuda', 'sync')
  geometric mean 0.020353968176404343
('exec_t', 'cuda', 'async')
  geometric mean 0.007937967178400238
('exec_t', 'x64', 'sync')
  geometric mean 0.0098667822306281
('exec_t', 'x64', 'async')
  geometric mean 0.004254368283956831
('launched_tasks', 'cuda', 'sync')
  geometric mean 36.660326883091024
('launched_tasks', 'cuda', 'async')
  geometric mean 9.819718193601348
('launched_tasks', 'x64', 'sync')
  geometric mean 31.281181322419652
('launched_tasks', 'x64', 'async')
  geometric mean 8.464726034771621
\fi

\subsection{MacCormack advection}
\label{sec:advection}

In this benchmark case, we use the MacCormack advection scheme~\cite{selle2008unconditionally} with RK3 path integration. We follow the recent trends to use collocated grids (see, e.g.,~\cite{nielsen2016spatially, gagniere2020hybrid}) to improve \changed{cacheline}{cache line} utilization. \revision{On a 3D unit grid, we advect three scalar fields of physically uncorrelated quantities over a sparsely populated vector field defined over a tube domain (see the supplemental document for more details). We use a multi-kernel implementation to keep the flexibility to use different schemes (stable fluid/MacCormack, possibly combined with decaying/sourcing) for different channels.}
\revision{Noticing the program is memory-bound, we expect an at most $3\times$ speedup when the three physical fields are advected together because of the improved cache line utilization when they are adjacent in memory. Our results show that our optimizer is able to achieve approximately $2.92\times$ and $2.80\times$ performance boost on CUDA and CPU, respectively, which is very close to the theoretical $3\times$ acceleration.} The number of launched tasks is reduced by $14.8\times$ on both backends. Compared to manually fused advection on different channels, our system automatically detects fusible patterns. This provides more coding flexibility and reduces the mental burden on developers.

We present an ablation study of four inter-kernel optimization passes in Table ~\ref{table:advection}. The improved performance and the reduced number of launched tasks in this benchmark mainly attribute to the task fusion and the list generation removal optimization.

\begin{table}[ht]
\caption{An ablation study on the MacCormack advection benchmark. The main optimization comes from task fusion. In this case, disabling list generation optimization transitively disables fusion, therefore degrades the performance. \reproduce{benchmark\_advect.py -d 3 -r 256 -n 100 -a [--no-lgr] [--no-ad] [--no-fusion] [--no-dse]}}
\scalebox{0.96}{
\begin{tabular}{lrr}
\toprule
\textbf{Ablation}   &  \textbf{Tasks} & \textbf{Wall-clock/GPU time (s)} \\ \midrule
\midrule
Reference~\cite{hu2019taichi} &   $4726$ & $8.498$/$8.479$ \\\midrule\midrule
No list generation removal    &   $3781$ & $8.542$/$8.488$ \\\midrule
No activation demotion        &   $319$ & $2.899$/$2.867$  \\\midrule
No task fusion                &   $950$ & $7.990$/$7.959$ \\\midrule 
No dead store elimination     &   $319$ & $2.906$/$2.869$ \\\midrule \midrule
All optimizations on          &   $319$ & $2.907$/$2.874$\\ \midrule
\bottomrule
\end{tabular}
}
\label{table:advection}
\end{table}

\subsection{Moving Least Squares Material Point Method (MLS-MPM)}
\label{sec:mlsmpm}

\begin{figure}[ht]
    \centering
    \begin{tabular}[t]{ccc}
        \hspace{-1mm}
        \begin{subfigure}{0.32\linewidth}
            \centering
            \includegraphics[width=\linewidth]{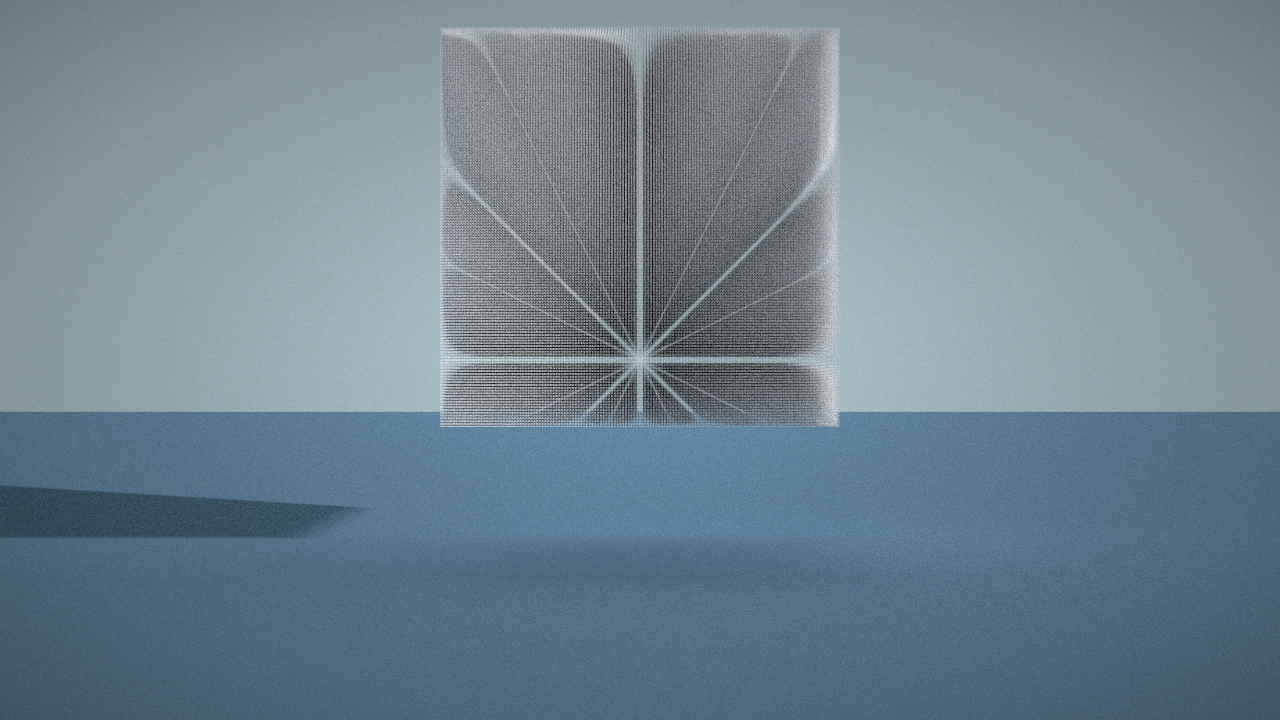}
        \end{subfigure}
        &
        \hspace{-4mm}
        \begin{subfigure}{0.32\linewidth}
            \centering
            \includegraphics[width=\linewidth]{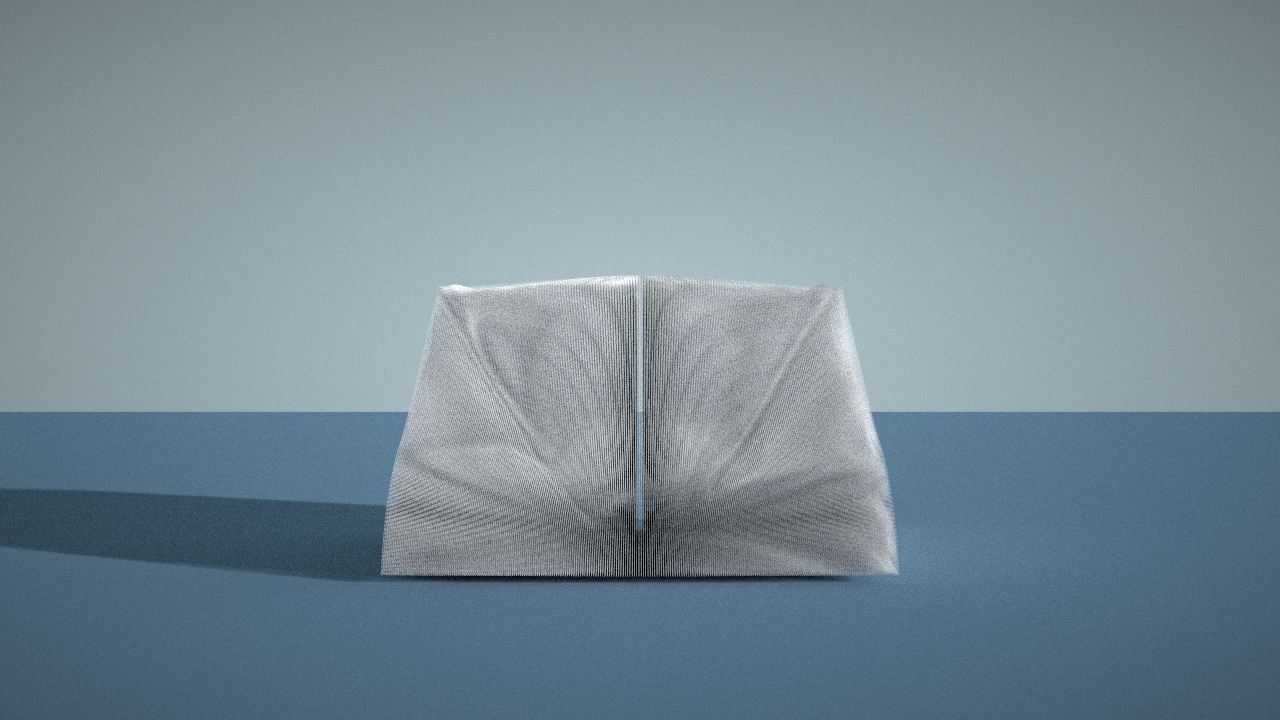}
        \end{subfigure}
        &
        \hspace{-4mm}
        \begin{subfigure}{0.32\linewidth}
            \centering
            \includegraphics[width=\linewidth]{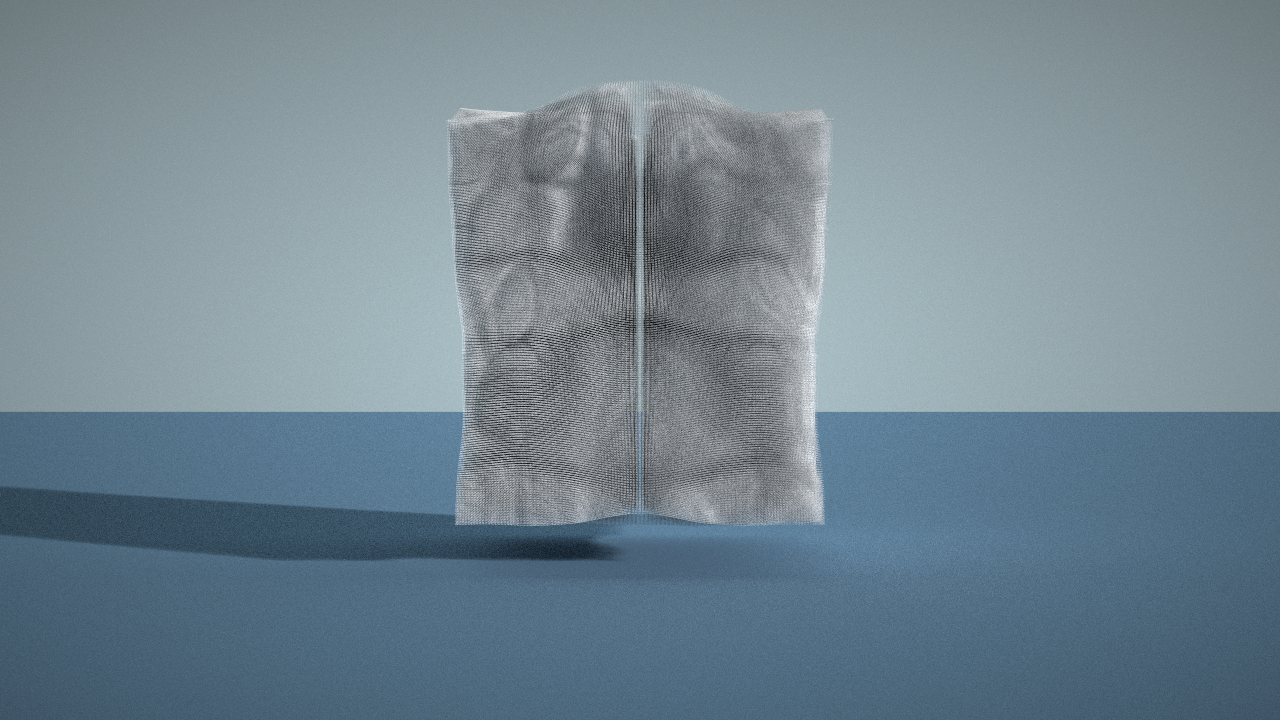}
        \end{subfigure}
	\end{tabular}
	\caption{The MLS-MPM benchmark case: a falling cube.}
	\label{fig:cube}
\end{figure}

\revision{Fig.~\ref{fig:cube} evaluates our system in more challenging cases where both particles and grids are used, where we benchmark against ~\cite{wang2020massively} (setup details are in the supplemental document).}

\vspace{-3mm}

\begin{figure}[h]
    \centering
    \includegraphics[width=0.99\linewidth]{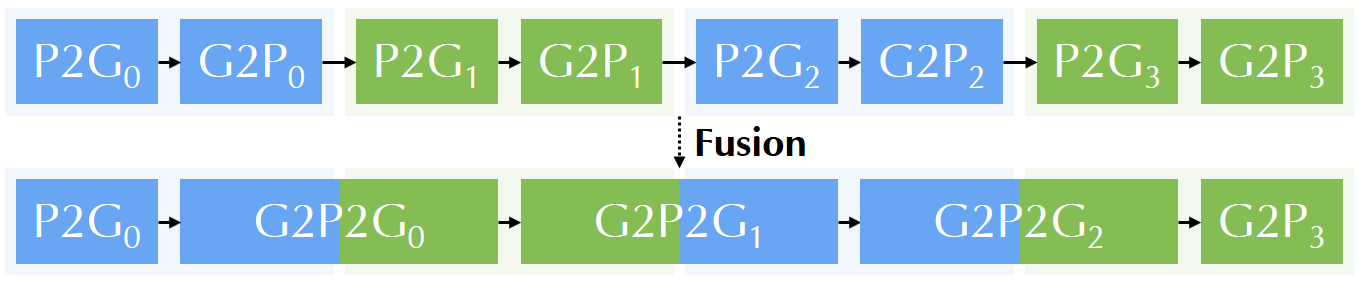}
    \caption{\revision{Our optimizer automatically fuses the consecutive G2P and P2G tasks. Writing P2G and G2P tasks separately is more modular and more common in practice, and fusing G2P with P2G in the next iteration leads to higher performance. Note that a functionally identical manual implementation of the fused version needs individual implementations of the P2G, G2P2G, and G2P kernels. Our automatic optimizer frees the programmers from doing tedious and error-prone code transforms, improving code performance at no cost of readability and maintainability.}}
    \label{fig:g2p2g}
\end{figure}

To achieve the best performance, we use two grids and swap them before each substep. \revision{We also hint our optimizer that the values stored in the \lstinline{pid} (particle ID) field is a permutation of all active indices in the \lstinline{particle} SNode (see the supplemental document for more details). Then our optimizer fuses G2P and P2G into a single G2P2G task (Fig.~\ref{fig:g2p2g}), which is the main source of optimization.}

\revision{We use a field \lstinline{C} to store the affine velocity around the particle (see~\cite{jiang2015affine}) between the G2P and P2G tasks. After fusing into a single G2P2G task, the dead store elimination pass further concludes that we do not need to store the intermediate result in the global field, which leads to further improved performance.} Our system achieves a $1.32\times$ performance boost on CUDA over the original Taichi system. The performance speedup due to fusion matches the originally reported number in ~\cite{wang2020massively}\footnote{Note that even in our implementation with G2P2G, Wang et al.~\cite{wang2020massively} (wall-clock time $5.004$ s on CUDA) is still $2.12\times$  faster than our system, because of their AOSOA acceleration data structure, which is outside the scope of this work. In their hand-engineered CUDA version, AOSOA+G2P2G is $2.1\times$ faster than G2P2G, which aligns well with the observations on our system.}.
\revision{We also benchmark against a manually fused G2P2G version implemented in the original Taichi system to mimic an optimized larger-scope megakernel with static fusion, and our system is $1.005\times$ faster than it on CUDA, indicating our automatic fusion system performs as fast as manual fusion.}

An ablation study of four inter-kernel optimizations is listed in Table ~\ref{table:mlsmpm}. In the microbenchmarks, there is a small-scale MPM test case (\lstinline{mpm_splitted}) that simply fuses per-particle operations and grid boundary conditions, leading to $1.1\times$ speed up.

\begin{table}[h]
\caption{An ablation study on the MLS-MPM benchmark. The main optimization comes from task fusion and dead store elimination. Note that without list generation removal, we cannot perform task fusion, and other optimizations will take much more time. \reproduce{benchmark\_mpm.py -n 2 -a [--no-lgr] [--no-ad] [--no-fusion] [--no-dse]}}
\scalebox{0.96}{
\begin{tabular}{lrr}
\toprule
\textbf{Ablation}   &  \textbf{Tasks} & \textbf{Wall-clock/GPU time (s)} \\ \midrule
\midrule
Reference~\cite{hu2019taichi} &   $18806$ & $14.059$/$13.987$ \\\midrule\midrule
No list generation removal    &   $14006$ & $15.059$/$13.908$ \\\midrule
No activation demotion        &   $9201$ & $10.976$/$10.927$  \\\midrule
No task fusion                &   $11608$ & $13.835$/$13.786$ \\\midrule 
No dead store elimination     &   $9201$ & $12.589$/$12.541$ \\\midrule \midrule
All optimizations on          &   $9201$ & $10.633$/$10.584$\\ \midrule
\bottomrule
\end{tabular}
}
\label{table:mlsmpm}
\end{table}

\subsection{Multigrid preconditioned conjugate gradients (MGPCG)}
\label{sec:mgpcg}

\begin{figure}[h]
    \centering
    \includegraphics[width=1.0\linewidth]{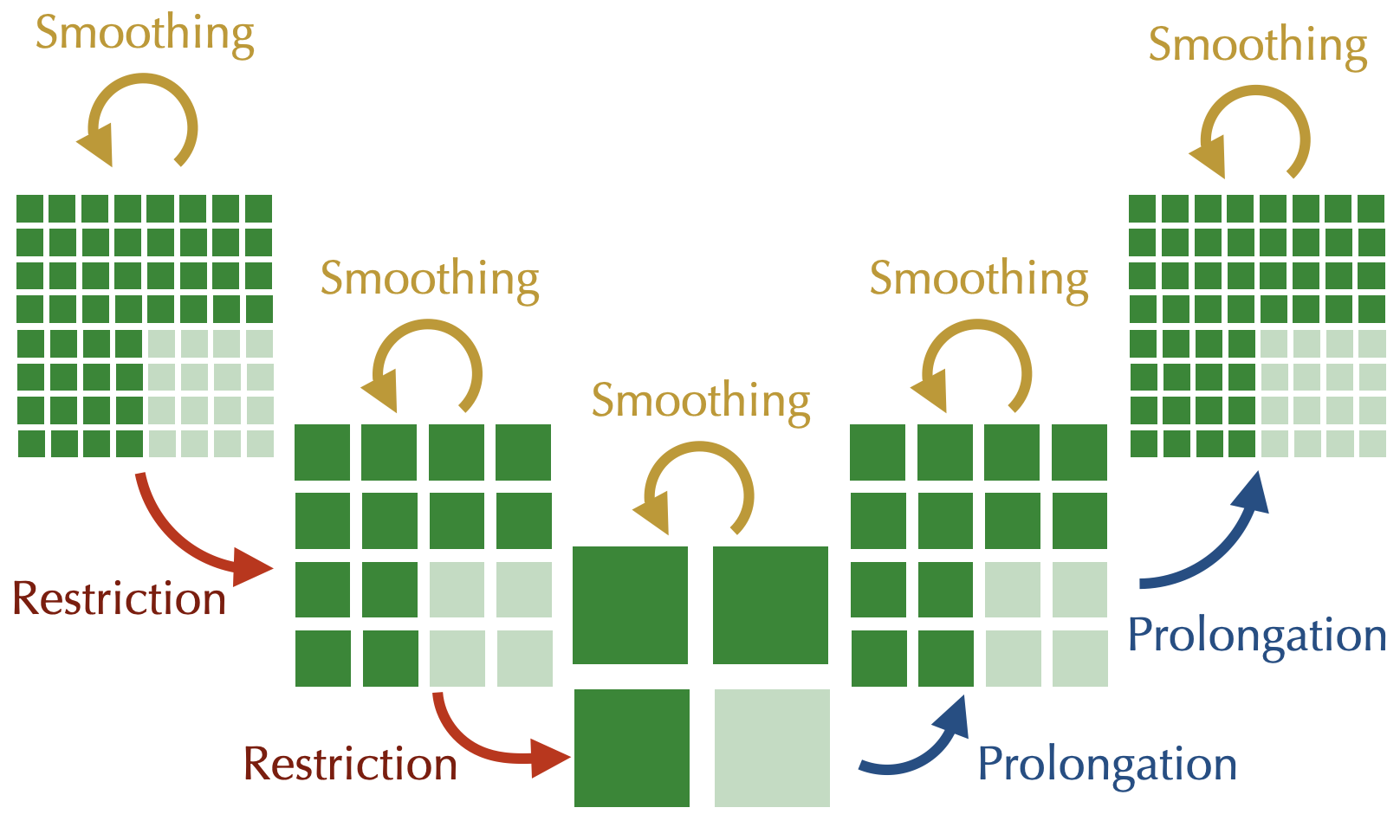}
    \caption{The key computational patterns in one iteration of the multigrid preconditioner. Note that in a single Poisson solve, once the sparse multigrid hierarchy is initialized, it will never change its topology. Our optimizer is able to infer the this property, since activation demotion will learn that the restriction kernels, starting from the second iteration, does not additionally activate any new voxels. The majority of list generations can also be removed.}
    \label{fig:mgpcg}
\end{figure}

Multigrid algorithms have frequent sparse data structure operations on grids of multiple resolutions. We test our system on a complex MGPCG Poisson solver. We use a sparsely populated region in a $1024\times 1024$ (2D) and $256\times 256 \times 256$ (3D) grid. We follow the MGPCG solver design in ~\cite{hu2019taichi}. In the 2D case, for example, our optimizer is able to bring down the number of tasks launched from $1,057,560$ to $224,568$ (only $21\%$ of the original number). This is because the restriction, smoothing, and prolongation operations lead to $420,912$ redundant list generation tasks, which are reduced to $8004$ ($2\%$ of the original) by our list generation removal and activation demotion (Fig.~\ref{fig:mgpcg}). Note that the CUDA speed ups ($2.49\times$ in 2D and $1.44\times$ in 3D) are much higher than the x64 speed up ($1.12\times$ in 2D and $1.07\times$ in 3D), likely because parallel task launches on GPUs are relatively more expensive than that on CPUs, and the majority of the speed ups in this benchmark case is from eliminating small kernels such as list generation and clearing.
\reproduce{benchmark\_mgpcg.py --arch [cpu/cuda] -d [2/3] -r [1024/256] -n 20}

We also evaluated the acceleration ratio of our optimizer, on problems of different scales. Note that part of the acceleration comes from reduced kernel launches, and larger-scale problems the kernel launching overhead is relatively smaller compared to more computation. Therefore, the wall-clock time improvement drops from $2.69\times$ to $2.23\times$ when the grid resolution raises from $128^2$ to $4096^2$. See our supplemental document for more details.

\begin{table}[h]
\caption{An ablation study on the 2D $1024\times 1024$ MGPCG benchmark. The main optimization is list generation removal. Note that activation demotion helps further eliminate $20\%$ of the list generation tasks \reproduce{benchmark\_mgpcg.py -d 2 -r 1024 -n 20 -a [--no-lgr] [--no-ad] [--no-fusion] [--no-dse]}}
\scalebox{0.97}{
\begin{tabular}{lrr}
\toprule
\textbf{Ablation}   &  \textbf{List / Total tasks} & \textbf{GPU time (s)} \\ \midrule
\midrule
Reference~\cite{hu2019taichi} &   $420912$/$1057560$ & $7.776$ \\\midrule\midrule
No list generation removal    &   $420912$/$827303$ & $7.232$ \\\midrule
No activation demotion        &   $10524$/$227104$ & $2.213$  \\\midrule
No task fusion                &   $8004$/$231695$ & $2.156$ \\\midrule 
No dead store elimination     &   $8004$/$224585$ & $2.098$ \\\midrule \midrule
All optimizations on          &   $8004$/$224565$ & $2.098$\\ \midrule
\bottomrule
\end{tabular}
}
\label{table:mgpcg}
\end{table}

\subsection{AutoDiff: nodal forces from energy gradients}
\label{sec:autodiff}

\begin{figure}[h]
    \centering
    \includegraphics[width=1.0\linewidth]{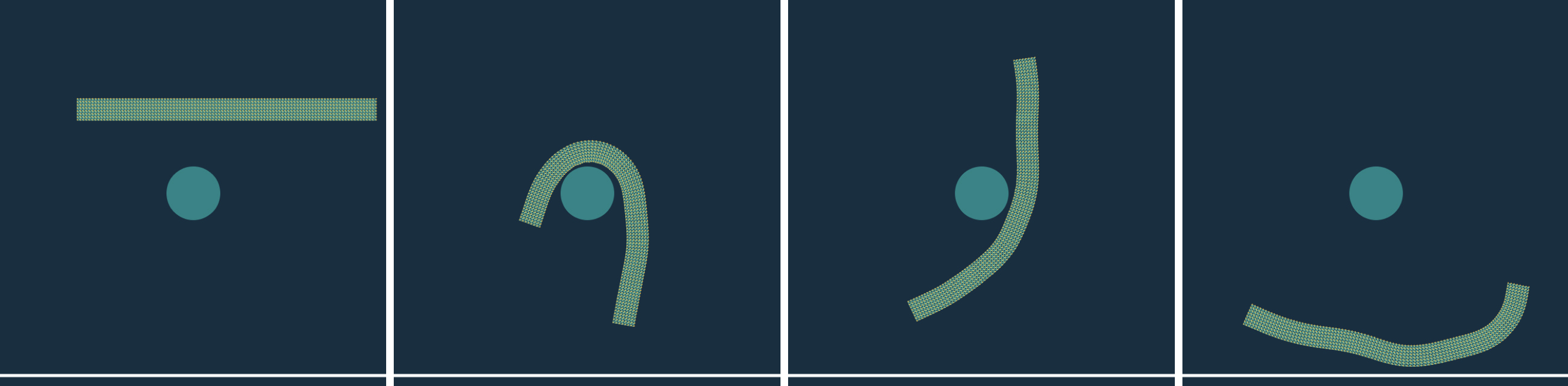}
    \caption{The AutoDiff test case: a hyper-elastic material falls down, hits an obstacle, and finally land on the ground. The nodal forces are computed using automatic differentiation. We use a low-resolution simulation here to better visualize the mesh structure.}
    \label{fig:autodiff}
\end{figure}

We implemented MLS-MPM~\cite{hu2018moving} with Lagrangian forces ~\cite{jiang2015affine}. In the simulation, the structure is modeled using a mesh of $160$K triangles, and a NeoHookean hyperelastic model (Fig.~\ref{fig:autodiff}). The force $\mathbf{f}_i$ on the particle $i$ is by definition
$$\mathbf{f}_i=-\frac{\partial L(\mathbf{x})}{\partial \mathbf{x}_i}.$$
\revision{where $L(\mathbf{x})$ is the Lagrangian.}

Since manually deriving the partial derivative on the right hand side is error-prone, we rely on Taichi's automatic differentiation system~\cite{hu2019difftaichi}. The key optimization opportunity is the following code:

\begin{lstlisting}
with ti.Tape(total_energy):
    compute_total_energy()
\end{lstlisting}

The code above does the forward computation of total energy $L(\mathbf{x})$, and then automatically evaluates for \lstinline{x.grad}, which is essentially $\frac{\partial L(\mathbf{x})}{\partial \mathbf{x}_i}$. \revision{In the majority of the cases where we only need the gradient rather than the total energy itself, the result of the total energy $L$ is \emph{not used}. By looking at a long window of kernels, our optimizer can automatically eliminate the forward computation, only doing the backward gradient evaluation.} Inter-kernel dead store elimination plays the most important role in this benchmark case.

An interesting observation is that our system gets \changed{}{a} significantly higher \changed{speed up}{speedup} on CUDA than x64. This is because the particle-to-grid (P2G) transfer step plays different roles in the total time consumption. Note that P2G requires atomic add, which is a relatively cheap operation on CUDA (native hardware support) yet expensive operation on x64 (needs software read-modify-write using hardware CAS). As a result, when our inter-kernel optimization is on, P2G takes $51\%$ run time on x64, yet only $7\%$ on CUDA. This means the forward total energy computation, which is optimized out, occupies \changed{}{a} smaller fraction on x64 (since P2G remains the bottleneck), hence a smaller speedup.

\begin{table}[h]
\caption{An ablation study on the AutoDiff benchmark. The main optimization comes from dead store elimination. Fusion helps\changed{}{,} to some extent, when evaluating and clearing the gradients. Since the data structures are all dense in this benchmark, list generation and activation demotion lead to no performance boost. \reproduce{benchmark\_autodiff.py -a [--no-lgr] [--no-ad] [--no-fusion] [--no-dse]}}
\scalebox{0.96}{
\begin{tabular}{lrr}
\toprule
\textbf{Ablation}   &  \textbf{Tasks} & \textbf{Wall-clock/GPU time (s)} \\ \midrule
\midrule
Reference~\cite{hu2019taichi} &   $65674$ & $2.256$/$2.214$ \\\midrule\midrule
No list generation removal    &   $42454$ & $1.386$/$1.191$ \\\midrule
No activation demotion        &   $42454$ & $1.383$/$1.188$  \\\midrule
No task fusion                &   $54064$ & $1.411$/$1.214$ \\\midrule 
No dead store elimination     &   $48424$ & $2.305$/$2.082$ \\\midrule \midrule
All optimizations on          &   $42454$ & $1.377$/$1.181$\\ \midrule
\bottomrule
\end{tabular}
}
\label{table:autodiff}
\end{table}

\subsection{Relationships to JAX}

\begin{figure}[h]
    \centering
    \includegraphics[width=0.9\linewidth]{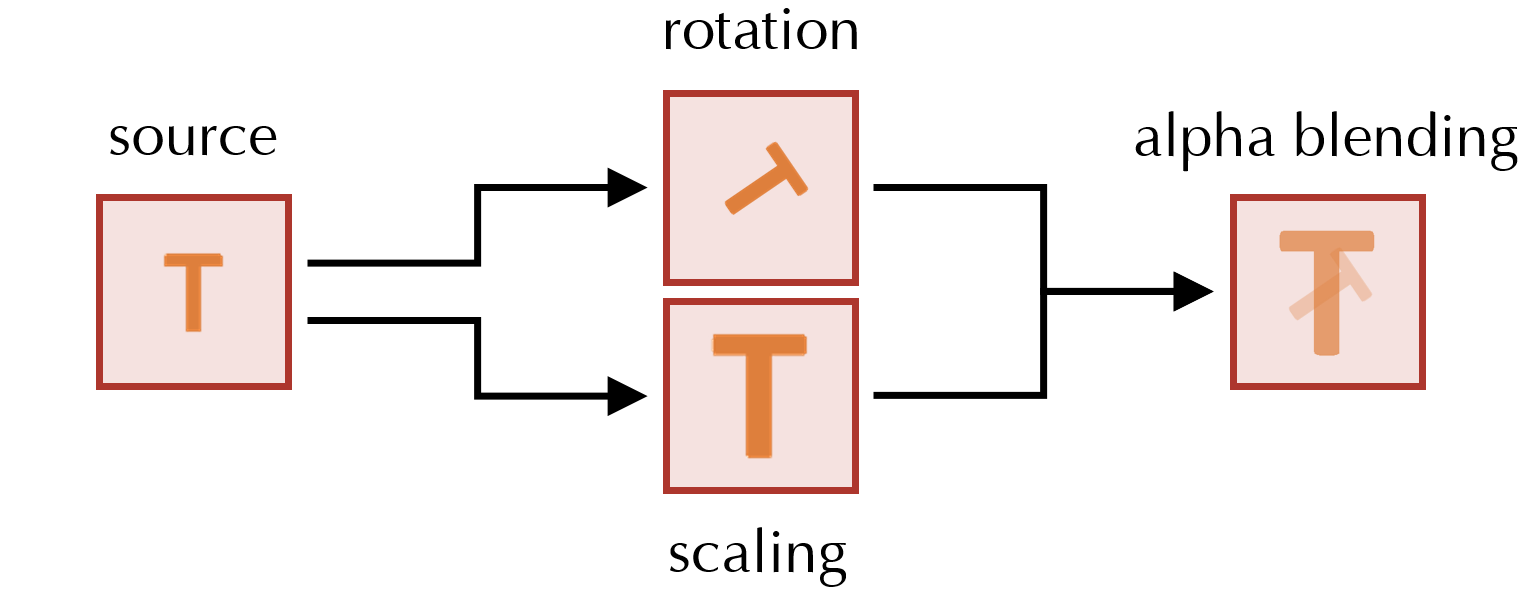}
    \caption{\revision{The texture generation test case: the source pattern is rotated and scaled separately using bilinear resampling, and then alpha blended into a final image.}}
    \label{fig:texture}
\end{figure}

\revision{Fusion is an effective optimization in both Taichi and JAX. The common high-level idea is to fuse CPU/GPU kernels to improve data locality and reduce kernel launches. However, due to different application domains and design decisions, fusion plays a more important role in JAX than in Taichi. Here we show a texture manipulation example (Fig.~\ref{fig:texture}) that lies at the intersection of application domains of Taichi and JAX.}

\begin{table}[h]
\caption{\revision{The GPU kernel execution time of the texture generation test case with/without fusion in Taichi and JAX. Taichi with fusion is our system, and Taichi without fusion is the original Taichi system~\cite{hu2019taichi}. JAX without fusion is JAX without the \lstinline{@jit} decorator. \reproducenvprof{benchmark\_texture\_gen.py [-a]; nvprof --print-gpu-trace python3 benchmark\_texture\_gen\_jax.py [--no-jit]}}}
\scalebox{0.97}{
\begin{tabular}{lrr}
\toprule
\textbf{Framework} & \textbf{with fusion (ms)} & \textbf{without fusion (ms)} \\
\midrule \midrule
Taichi & 0.306 & 0.467 \\\midrule
JAX & 1.005 & 24.462 \\ \midrule
\bottomrule
\end{tabular}
}
\label{table:texture}
\end{table}

\revision{We implemented the program using both Taichi and JAX. It turns out that Taichi without fusion, Taichi is $52.4\times$ faster than JAX, but with fusion Taichi is only $3.3\times$ faster than JAX. We implemented each step using a kernel in Taichi, and operations in a Taichi kernel are already ``fused" by default. This leads to a higher arithmetic intensity (i.e., FLOPs per byte fetched from memory) and fewer kernel launches in Taichi than in JAX when fusion is disabled, hence the significant performance advantage. High-level relationships between fusion in Taichi and that in JAX are listed below:}

\revision{
\begin{itemize}
    \item \textbf{Efficiency improvement.} Fusion in JAX creates a greater performance boost than that in Taichi. This is because a single Taichi (mega)kernel can already be considered as a fused version of underlying operators. Still, further fusing these kernels in Taichi helps to some extent. 
    \item \textbf{Analysis difficulty.} Fusion in JAX is easier compared to that in Taichi. This is because \textbf{1)} fusion in JAX works only on pure functions and in-place mutating updates of arrays are not supported~\footnote{More details on the  \textcolor{blue}{\href{https://github.com/google/jax/tree/25cc3ece66879c4cf154c8382fe59083d9cabaa9\#current-gotchas}{JAX GitHub}}.}, and \textbf{2)} most JAX operations are tensor expressions that take holistic arrays as inputs and outputs (e.g., $a=b+c$) while Taichi kernels are often finer-grained array manipulations (e.g., $a_{i,j-i}+\!=b_{i,j+3}+c_{j, 2i}$). These two reasons make it easy to model the computational graph using a feed-forward DAG, without the need for relatively more complex data-flow modeling as in Section~\ref{sec:sfg}.
\end{itemize}
}

\subsection{Discussions}

\label{sec:timeline}

\paragraph{Productivity} An attractive feature of our system is users get \changed{}{a} performance boost for {\em free}, simply by turning on an option to let the system conduct inter-kernel optimizations.

\begin{figure*}[h]
    \centering
    \includegraphics[width=1.0\linewidth]{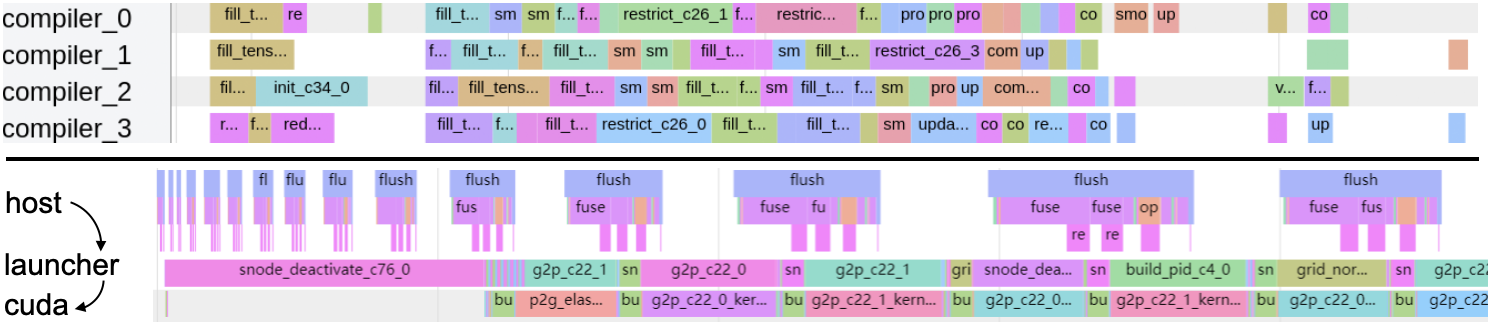}
    \caption{\textbf{Top:} In our asynchronous execution engine, multiple compilation threads simultaneously compile optimized IR to executable kernels, maximizing JIT benefits. Timelines are gathered from the MGPCG example, where our parallel compilation system leads to a $2.43\times$ shorted program start-up time. \textbf{Bottom:} At later stages of program execution, most possible inter-kernel optimized kernels are already compiled and cached, so the compilation threads are mostly idle. Still, our multithreading framework allows the inter-kernel optimizer to overlap with the launcher thread and GPUs, making sure no GPU starvation happens. Timelines are gathered from the MLS-MPM example.}
    \label{fig:timeline}
\end{figure*}

\paragraph{Parallel compilation} 
Delay caused by compilation time may lead to potential performance issues in JIT systems. Fortunately, in our asynchronous execution engine\changed{}{,} we have a whole buffer of kernels to compile, and parallel compilation significantly reduces this compilation delay. For example, in the 2D $1024\times 1024$ MGPCG benchmark, we find the time of the first iteration, which is compilation delay as no kernels are compiled and cached, improves from $9.0s$ to $3.7s$ after switching to the asynchronous engine, a $2.43\times$ improvement. See Fig.~\ref{fig:timeline} for a visualization of the multithreading behavior of the asynchronous optimization and execution engine.

\paragraph{Behavior on CPUs}
Interestingly, on CPU\changed{}{,} the performance boost is less significant compared to GPUs. Initial investigations show three reasons:
\begin{enumerate}
    \item When using the CPU backend\changed{}{,} the optimizer and executor share the same processor, meaning the optimization process itself may slow down the execution. This issue does not exist on the GPU backend, since optimization overlaps with the GPU kernel execution time.
    \item On CPU computation itself occupies a bigger fraction of the execution time. For example, in the AutoDiff example\changed{}{,} we find $51$\% of \changed{}{the} execution time was spent on scattering force contributions to nodes, which needs expensive software-emulated atomic add. On GPUs\changed{}{,} atomic add is hardware-native and is much faster. When the task is fully memory-bound, e.g., the MacCormack advection benchmark, we do achieve a close to ideal $3\times$ speed up on CPUs, similar to the behavior on GPUs.
    \item Some of our performance boost comes from eliminating kernels. Compared to the actual computation, kernel launching is expensive on GPU but relatively cheap on CPU.
\end{enumerate}

\paragraph{Wall-clock time v.s. backend time} In most cases\changed{}{,} our wall-clock time is within $103\%$ of the execution time, which indicates that our multi-threaded optimization and compilation system is able to keep GPU busy. However, in the 2D MGPCG example, we find wall-clock time to be $1.75\times$ higher than the execution time. This is an engineering limitation of our work: some optimization passes, such as kernel fusion, still takes \changed{}{a} longer time than expected on relatively small-scale problems with many tasks. We believe a more carefully engineered optimization system can get rid of this issue.

\section{Conclusion}
\revision{We have presented an inter-kernel optimization system for parallel, imperative, and spatially sparse computation. In our benchmark cases, we achieved an $1.87\times$ wall-clock time performance improvement, without users changing any computation code. We believe our system can alleviate the low-level performance engineering burden on programmers, achieving higher-performance spatially sparse computation without harming code modularity and readability.}

\paragraph{Future work}

\revision{While our optimizer improves the runtime performance of spatially sparse computation, it does not automatically eliminate the storage cost of intermediate fields. This is partly due to a limitation of Taichi: it materializes the entire data structure before executing any kernel, then no fields can be created or destroyed. To address this problem, Taichi needs a system extension to support dynamic creation and destruction of fields. A tailored optimizer can then be designed.}

\revision{Meanwhile, there are many unexplored kernel metadata we can extract for sparse computation. For example, a colored Gauss-Seidel solver may only use the ``white'' cells in a checkerboard pattern. These features may enable further inter-kernel optimization opportunities in physical simulation and numerical linear algebra.}

\bibliographystyle{ACM-Reference-Format}
\bibliography{ref}


\begin{thebibliography}{39}


\ifx \showCODEN    \undefined \def \showCODEN     #1{\unskip}     \fi
\ifx \showDOI      \undefined \def \showDOI       #1{#1}\fi
\ifx \showISBNx    \undefined \def \showISBNx     #1{\unskip}     \fi
\ifx \showISBNxiii \undefined \def \showISBNxiii  #1{\unskip}     \fi
\ifx \showISSN     \undefined \def \showISSN      #1{\unskip}     \fi
\ifx \showLCCN     \undefined \def \showLCCN      #1{\unskip}     \fi
\ifx \shownote     \undefined \def \shownote      #1{#1}          \fi
\ifx \showarticletitle \undefined \def \showarticletitle #1{#1}   \fi
\ifx \showURL      \undefined \def \showURL       {\relax}        \fi
\providecommand\bibfield[2]{#2}
\providecommand\bibinfo[2]{#2}
\providecommand\natexlab[1]{#1}
\providecommand\showeprint[2][]{arXiv:#2}

\bibitem[\protect\citeauthoryear{Abadi, Barham, Chen, Chen, Davis, Dean, Devin,
  Ghemawat, Irving, Isard, Kudlur, Levenberg, Monga, Moore, Murray, Steiner,
  Tucker, Vasudevan, Warden, Wicke, Yu, and Zheng}{Abadi et~al\mbox{.}}{2016}]%
        {45381}
\bibfield{author}{\bibinfo{person}{Martin Abadi}, \bibinfo{person}{Paul
  Barham}, \bibinfo{person}{Jianmin Chen}, \bibinfo{person}{Zhifeng Chen},
  \bibinfo{person}{Andy Davis}, \bibinfo{person}{Jeffrey Dean},
  \bibinfo{person}{Matthieu Devin}, \bibinfo{person}{Sanjay Ghemawat},
  \bibinfo{person}{Geoffrey Irving}, \bibinfo{person}{Michael Isard},
  \bibinfo{person}{Manjunath Kudlur}, \bibinfo{person}{Josh Levenberg},
  \bibinfo{person}{Rajat Monga}, \bibinfo{person}{Sherry Moore},
  \bibinfo{person}{Derek~G. Murray}, \bibinfo{person}{Benoit Steiner},
  \bibinfo{person}{Paul Tucker}, \bibinfo{person}{Vijay Vasudevan},
  \bibinfo{person}{Pete Warden}, \bibinfo{person}{Martin Wicke},
  \bibinfo{person}{Yuan Yu}, {and} \bibinfo{person}{Xiaoqiang Zheng}.}
  \bibinfo{year}{2016}\natexlab{}.
\newblock \showarticletitle{TensorFlow: A system for large-scale machine
  learning}. In \bibinfo{booktitle}{\emph{12th USENIX Symposium on Operating
  Systems Design and Implementation (OSDI 16)}}. \bibinfo{pages}{265--283}.
\newblock
\urldef\tempurl%
\url{https://www.usenix.org/system/files/conference/osdi16/osdi16-abadi.pdf}
\showURL{%
\tempurl}


\bibitem[\protect\citeauthoryear{Bai, Lu, Zhang, et~al\mbox{.}}{Bai
  et~al\mbox{.}}{2019}]%
        {bai2019}
\bibfield{author}{\bibinfo{person}{Junjie Bai}, \bibinfo{person}{Fang Lu},
  \bibinfo{person}{Ke Zhang}, {et~al\mbox{.}}} \bibinfo{year}{2019}\natexlab{}.
\newblock \bibinfo{title}{ONNX: Open Neural Network Exchange}.
\newblock \bibinfo{howpublished}{\url{https://github.com/onnx/onnx}}.
\newblock


\bibitem[\protect\citeauthoryear{Bernstein, Shah, Lemire, Devito, Fisher,
  Levis, and Hanrahan}{Bernstein et~al\mbox{.}}{2016}]%
        {Bernstein:2016:EDP}
\bibfield{author}{\bibinfo{person}{Gilbert~Louis Bernstein},
  \bibinfo{person}{Chinmayee Shah}, \bibinfo{person}{Crystal Lemire},
  \bibinfo{person}{Zachary Devito}, \bibinfo{person}{Matthew Fisher},
  \bibinfo{person}{Philip Levis}, {and} \bibinfo{person}{Pat Hanrahan}.}
  \bibinfo{year}{2016}\natexlab{}.
\newblock \showarticletitle{Ebb: A {DSL} for physical simulation on {CPUs} and
  {GPUs}}.
\newblock \bibinfo{journal}{\emph{ACM Trans. Graph.}} \bibinfo{volume}{35},
  \bibinfo{number}{2} (\bibinfo{year}{2016}), \bibinfo{pages}{21:1--21:12}.
\newblock


\bibitem[\protect\citeauthoryear{Bradbury, Frostig, Hawkins, Johnson, Leary,
  Maclaurin, Necula, Paszke, Vander{P}las, Wanderman-{M}ilne, and
  Zhang}{Bradbury et~al\mbox{.}}{2018}]%
        {jax2018github}
\bibfield{author}{\bibinfo{person}{James Bradbury}, \bibinfo{person}{Roy
  Frostig}, \bibinfo{person}{Peter Hawkins}, \bibinfo{person}{Matthew~James
  Johnson}, \bibinfo{person}{Chris Leary}, \bibinfo{person}{Dougal Maclaurin},
  \bibinfo{person}{George Necula}, \bibinfo{person}{Adam Paszke},
  \bibinfo{person}{Jake Vander{P}las}, \bibinfo{person}{Skye
  Wanderman-{M}ilne}, {and} \bibinfo{person}{Qiao Zhang}.}
  \bibinfo{year}{2018}\natexlab{}.
\newblock \bibinfo{title}{{JAX}: composable transformations of
  {P}ython+{N}um{P}y programs}.
\newblock
\newblock
\urldef\tempurl%
\url{http://github.com/google/jax}
\showURL{%
\tempurl}


\bibitem[\protect\citeauthoryear{Briggs, Evans, Grant, Hundt, Maddox, Novillo,
  Park, Sehr, Taylor, and Wild}{Briggs et~al\mbox{.}}{2007}]%
        {briggs2007whopr}
\bibfield{author}{\bibinfo{person}{Preston Briggs}, \bibinfo{person}{Doug
  Evans}, \bibinfo{person}{Brian Grant}, \bibinfo{person}{Robert Hundt},
  \bibinfo{person}{William Maddox}, \bibinfo{person}{Diego Novillo},
  \bibinfo{person}{Seongbae Park}, \bibinfo{person}{David Sehr},
  \bibinfo{person}{Ian Taylor}, {and} \bibinfo{person}{Ollie Wild}.}
  \bibinfo{year}{2007}\natexlab{}.
\newblock \showarticletitle{WHOPR-fast and scalable whole program optimizations
  in GCC}.
\newblock \bibinfo{journal}{\emph{Initial Draft}}  \bibinfo{volume}{12}
  (\bibinfo{year}{2007}).
\newblock


\bibitem[\protect\citeauthoryear{Chou, Kjolstad, and Amarasinghe}{Chou
  et~al\mbox{.}}{2018}]%
        {chou2018format}
\bibfield{author}{\bibinfo{person}{Stephen Chou}, \bibinfo{person}{Fredrik
  Kjolstad}, {and} \bibinfo{person}{Saman Amarasinghe}.}
  \bibinfo{year}{2018}\natexlab{}.
\newblock \showarticletitle{Format abstraction for sparse tensor algebra
  compilers}.
\newblock \bibinfo{journal}{\emph{Proceedings of the ACM on Programming
  Languages}} \bibinfo{volume}{2}, \bibinfo{number}{OOPSLA}
  (\bibinfo{year}{2018}), \bibinfo{pages}{1--30}.
\newblock


\bibitem[\protect\citeauthoryear{DeVito, Joubert, Palacios, Oakley, Medina,
  Barrientos, Elsen, Ham, Aiken, Duraisamy, et~al\mbox{.}}{DeVito
  et~al\mbox{.}}{2011}]%
        {Devito:2011:LDS}
\bibfield{author}{\bibinfo{person}{Zachary DeVito}, \bibinfo{person}{Niels
  Joubert}, \bibinfo{person}{Francisco Palacios}, \bibinfo{person}{Stephen
  Oakley}, \bibinfo{person}{Montserrat Medina}, \bibinfo{person}{Mike
  Barrientos}, \bibinfo{person}{Erich Elsen}, \bibinfo{person}{Frank Ham},
  \bibinfo{person}{Alex Aiken}, \bibinfo{person}{Karthik Duraisamy},
  {et~al\mbox{.}}} \bibinfo{year}{2011}\natexlab{}.
\newblock \showarticletitle{Liszt: A domain specific language for building
  portable mesh-based PDE solvers}. In \bibinfo{booktitle}{\emph{International
  Conference for High Performance Computing, Networking, Storage and
  Analysis}}. \bibinfo{pages}{9}.
\newblock


\bibitem[\protect\citeauthoryear{Filipovi{\v{c}}, Madzin, Fousek, and
  Matyska}{Filipovi{\v{c}} et~al\mbox{.}}{2015}]%
        {filipovivc2015optimizing}
\bibfield{author}{\bibinfo{person}{Ji{\v{r}}{\'\i} Filipovi{\v{c}}},
  \bibinfo{person}{Mat{\'u}{\v{s}} Madzin}, \bibinfo{person}{Jan Fousek}, {and}
  \bibinfo{person}{Lud{\v{e}}k Matyska}.} \bibinfo{year}{2015}\natexlab{}.
\newblock \showarticletitle{Optimizing CUDA code by kernel fusion: application
  on BLAS}.
\newblock \bibinfo{journal}{\emph{The Journal of Supercomputing}}
  \bibinfo{volume}{71}, \bibinfo{number}{10} (\bibinfo{year}{2015}),
  \bibinfo{pages}{3934--3957}.
\newblock


\bibitem[\protect\citeauthoryear{Gagniere, Hyde, Marquez-Razon, Jiang, Ge, Han,
  Guo, and Teran}{Gagniere et~al\mbox{.}}{2020}]%
        {gagniere2020hybrid}
\bibfield{author}{\bibinfo{person}{Steven~W Gagniere},
  \bibinfo{person}{David~AB Hyde}, \bibinfo{person}{Alan Marquez-Razon},
  \bibinfo{person}{Chenfanfu Jiang}, \bibinfo{person}{Ziheng Ge},
  \bibinfo{person}{Xuchen Han}, \bibinfo{person}{Qi Guo}, {and}
  \bibinfo{person}{Joseph Teran}.} \bibinfo{year}{2020}\natexlab{}.
\newblock \showarticletitle{A Hybrid Lagrangian/Eulerian Collocated Advection
  and Projection Method for Fluid Simulation}.
\newblock \bibinfo{journal}{\emph{arXiv preprint arXiv:2003.12227}}
  (\bibinfo{year}{2020}).
\newblock


\bibitem[\protect\citeauthoryear{Gao, Wang, Wu, Pradhana-Tampubolon, Sifakis,
  Cem, and Jiang}{Gao et~al\mbox{.}}{2018}]%
        {gao2018gpu}
\bibfield{author}{\bibinfo{person}{Ming Gao}, \bibinfo{person}{Xinlei Wang},
  \bibinfo{person}{Kui Wu}, \bibinfo{person}{Andre Pradhana-Tampubolon},
  \bibinfo{person}{Eftychios Sifakis}, \bibinfo{person}{Yuksel Cem}, {and}
  \bibinfo{person}{Chenfanfu Jiang}.} \bibinfo{year}{2018}\natexlab{}.
\newblock \showarticletitle{{GPU} Optimization of Material Point Methods}.
\newblock \bibinfo{journal}{\emph{ACM Trans. Graph. (Proc. SIGGRAPH Asia)}}
  \bibinfo{volume}{32}, \bibinfo{number}{4} (\bibinfo{year}{2018}),
  \bibinfo{pages}{102}.
\newblock


\bibitem[\protect\citeauthoryear{Hoetzlein}{Hoetzlein}{2016}]%
        {hoetzlein2016gvdb}
\bibfield{author}{\bibinfo{person}{Rama~Karl Hoetzlein}.}
  \bibinfo{year}{2016}\natexlab{}.
\newblock \showarticletitle{GVDB: Raytracing sparse voxel database structures
  on the {GPU}}. In \bibinfo{booktitle}{\emph{Proceedings of High Performance
  Graphics}}. Eurographics Association, \bibinfo{pages}{109--117}.
\newblock


\bibitem[\protect\citeauthoryear{Hong, Sukumaran-Rajam, Kim, Rawat,
  Krishnamoorthy, Pouchet, Rastello, and Sadayappan}{Hong
  et~al\mbox{.}}{2018}]%
        {hong2018gpu}
\bibfield{author}{\bibinfo{person}{Changwan Hong}, \bibinfo{person}{Aravind
  Sukumaran-Rajam}, \bibinfo{person}{Jinsung Kim},
  \bibinfo{person}{Prashant~Singh Rawat}, \bibinfo{person}{Sriram
  Krishnamoorthy}, \bibinfo{person}{Louis-No{\"e}l Pouchet},
  \bibinfo{person}{Fabrice Rastello}, {and} \bibinfo{person}{P Sadayappan}.}
  \bibinfo{year}{2018}\natexlab{}.
\newblock \showarticletitle{Gpu code optimization using abstract kernel
  emulation and sensitivity analysis}. In \bibinfo{booktitle}{\emph{Proceedings
  of the 39th ACM SIGPLAN Conference on Programming Language Design and
  Implementation}}. \bibinfo{pages}{736--751}.
\newblock


\bibitem[\protect\citeauthoryear{Hu}{Hu}{2020}]%
        {hu2020taichi}
\bibfield{author}{\bibinfo{person}{Yuanming Hu}.}
  \bibinfo{year}{2020}\natexlab{}.
\newblock \showarticletitle{The Taichi programming language}.
\newblock In \bibinfo{booktitle}{\emph{ACM SIGGRAPH 2020 Courses}}.
  \bibinfo{pages}{1--50}.
\newblock


\bibitem[\protect\citeauthoryear{Hu, Anderson, Li, Sun, Carr, Ragan-Kelley, and
  Durand}{Hu et~al\mbox{.}}{2020}]%
        {hu2019difftaichi}
\bibfield{author}{\bibinfo{person}{Yuanming Hu}, \bibinfo{person}{Luke
  Anderson}, \bibinfo{person}{Tzu-Mao Li}, \bibinfo{person}{Qi Sun},
  \bibinfo{person}{Nathan Carr}, \bibinfo{person}{Jonathan Ragan-Kelley}, {and}
  \bibinfo{person}{Fr{\'e}do Durand}.} \bibinfo{year}{2020}\natexlab{}.
\newblock \showarticletitle{DiffTaichi: Differentiable Programming for Physical
  Simulation}.
\newblock \bibinfo{journal}{\emph{ICLR}} (\bibinfo{year}{2020}).
\newblock


\bibitem[\protect\citeauthoryear{Hu, Fang, Ge, Qu, Zhu, Pradhana, and Jiang}{Hu
  et~al\mbox{.}}{2018}]%
        {hu2018moving}
\bibfield{author}{\bibinfo{person}{Yuanming Hu}, \bibinfo{person}{Yu Fang},
  \bibinfo{person}{Ziheng Ge}, \bibinfo{person}{Ziyin Qu},
  \bibinfo{person}{Yixin Zhu}, \bibinfo{person}{Andre Pradhana}, {and}
  \bibinfo{person}{Chenfanfu Jiang}.} \bibinfo{year}{2018}\natexlab{}.
\newblock \showarticletitle{A moving least squares material point method with
  displacement discontinuity and two-way rigid body coupling}.
\newblock \bibinfo{journal}{\emph{ACM Trans. Graph. (Proc. SIGGRAPH Asia)}}
  \bibinfo{volume}{37}, \bibinfo{number}{4} (\bibinfo{year}{2018}),
  \bibinfo{pages}{150}.
\newblock


\bibitem[\protect\citeauthoryear{Hu, Li, Anderson, Ragan-Kelley, and Durand}{Hu
  et~al\mbox{.}}{2019}]%
        {hu2019taichi}
\bibfield{author}{\bibinfo{person}{Yuanming Hu}, \bibinfo{person}{Tzu-Mao Li},
  \bibinfo{person}{Luke Anderson}, \bibinfo{person}{Jonathan Ragan-Kelley},
  {and} \bibinfo{person}{Fr{\'e}do Durand}.} \bibinfo{year}{2019}\natexlab{}.
\newblock \showarticletitle{Taichi: a language for high-performance computation
  on spatially sparse data structures}.
\newblock \bibinfo{journal}{\emph{ACM Transactions on Graphics (TOG)}}
  \bibinfo{volume}{38}, \bibinfo{number}{6} (\bibinfo{year}{2019}),
  \bibinfo{pages}{201}.
\newblock


\bibitem[\protect\citeauthoryear{Hu, Liu, Yang, Xu, Kuang, Xu, Dai, Freeman,
  and Durand}{Hu et~al\mbox{.}}{2021}]%
        {hu2021quantaichi}
\bibfield{author}{\bibinfo{person}{Yuanming Hu}, \bibinfo{person}{Jiafeng Liu},
  \bibinfo{person}{Xuanda Yang}, \bibinfo{person}{Mingkuan Xu},
  \bibinfo{person}{Ye Kuang}, \bibinfo{person}{Weiwei Xu},
  \bibinfo{person}{Qiang Dai}, \bibinfo{person}{William~T. Freeman}, {and}
  \bibinfo{person}{Frédo Durand}.} \bibinfo{year}{2021}\natexlab{}.
\newblock \showarticletitle{QuanTaichi: A Compiler for Quantized Simulations}.
\newblock \bibinfo{journal}{\emph{ACM Transactions on Graphics (TOG)}}
  \bibinfo{volume}{40}, \bibinfo{number}{4} (\bibinfo{year}{2021}).
\newblock


\bibitem[\protect\citeauthoryear{Jiang, Schroeder, Selle, Teran, and
  Stomakhin}{Jiang et~al\mbox{.}}{2015}]%
        {jiang2015affine}
\bibfield{author}{\bibinfo{person}{Chenfanfu Jiang}, \bibinfo{person}{Craig
  Schroeder}, \bibinfo{person}{Andrew Selle}, \bibinfo{person}{Joseph Teran},
  {and} \bibinfo{person}{Alexey Stomakhin}.} \bibinfo{year}{2015}\natexlab{}.
\newblock \showarticletitle{The affine particle-in-cell method}.
\newblock \bibinfo{journal}{\emph{ACM Transactions on Graphics (TOG)}}
  \bibinfo{volume}{34}, \bibinfo{number}{4} (\bibinfo{year}{2015}),
  \bibinfo{pages}{1--10}.
\newblock


\bibitem[\protect\citeauthoryear{Khedker, Sanyal, and Sathe}{Khedker
  et~al\mbox{.}}{2017}]%
        {khedker2017data}
\bibfield{author}{\bibinfo{person}{Uday Khedker}, \bibinfo{person}{Amitabha
  Sanyal}, {and} \bibinfo{person}{Bageshri Sathe}.}
  \bibinfo{year}{2017}\natexlab{}.
\newblock \bibinfo{booktitle}{\emph{Data flow analysis: theory and practice}}.
\newblock \bibinfo{publisher}{CRC Press}.
\newblock


\bibitem[\protect\citeauthoryear{Kjolstad, Kamil, Chou, Lugato, and
  Amarasinghe}{Kjolstad et~al\mbox{.}}{2017}]%
        {kjolstad2017tensor}
\bibfield{author}{\bibinfo{person}{Fredrik Kjolstad}, \bibinfo{person}{Shoaib
  Kamil}, \bibinfo{person}{Stephen Chou}, \bibinfo{person}{David Lugato}, {and}
  \bibinfo{person}{Saman Amarasinghe}.} \bibinfo{year}{2017}\natexlab{}.
\newblock \showarticletitle{The tensor algebra compiler}.
\newblock \bibinfo{journal}{\emph{Proceedings of the ACM on Programming
  Languages}} \bibinfo{volume}{1}, \bibinfo{number}{OOPSLA}
  (\bibinfo{year}{2017}), \bibinfo{pages}{1--29}.
\newblock


\bibitem[\protect\citeauthoryear{Kjolstad, Kamil, Ragan-Kelley, Levin, Sueda,
  Chen, Vouga, Kaufman, Kanwar, Matusik, and Amarasinghe}{Kjolstad
  et~al\mbox{.}}{2016}]%
        {Kjolstad:2016:SLP}
\bibfield{author}{\bibinfo{person}{Fredrik Kjolstad}, \bibinfo{person}{Shoaib
  Kamil}, \bibinfo{person}{Jonathan Ragan-Kelley}, \bibinfo{person}{David I.~W.
  Levin}, \bibinfo{person}{Shinjiro Sueda}, \bibinfo{person}{Desai Chen},
  \bibinfo{person}{Etienne Vouga}, \bibinfo{person}{Danny~M. Kaufman},
  \bibinfo{person}{Gurtej Kanwar}, \bibinfo{person}{Wojciech Matusik}, {and}
  \bibinfo{person}{Saman Amarasinghe}.} \bibinfo{year}{2016}\natexlab{}.
\newblock \showarticletitle{Simit: A language for physical simulation}.
\newblock \bibinfo{journal}{\emph{ACM Trans. Graph.}} \bibinfo{volume}{35},
  \bibinfo{number}{2} (\bibinfo{year}{2016}), \bibinfo{pages}{20:1--20:21}.
\newblock


\bibitem[\protect\citeauthoryear{Knobe and Sarkar}{Knobe and Sarkar}{1998}]%
        {knobe1998array}
\bibfield{author}{\bibinfo{person}{Kathleen Knobe} {and} \bibinfo{person}{Vivek
  Sarkar}.} \bibinfo{year}{1998}\natexlab{}.
\newblock \showarticletitle{Array SSA form and its use in parallelization}. In
  \bibinfo{booktitle}{\emph{Proceedings of the 25th ACM SIGPLAN-SIGACT
  symposium on Principles of programming languages}}.
  \bibinfo{pages}{107--120}.
\newblock


\bibitem[\protect\citeauthoryear{{Lattner}, {Amini}, {Bondhugula}, {Cohen},
  {Davis}, {Pienaar}, {Riddle}, {Shpeisman}, {Vasilache}, and
  {Zinenko}}{{Lattner} et~al\mbox{.}}{2021}]%
        {9370308}
\bibfield{author}{\bibinfo{person}{C. {Lattner}}, \bibinfo{person}{M. {Amini}},
  \bibinfo{person}{U. {Bondhugula}}, \bibinfo{person}{A. {Cohen}},
  \bibinfo{person}{A. {Davis}}, \bibinfo{person}{J. {Pienaar}},
  \bibinfo{person}{R. {Riddle}}, \bibinfo{person}{T. {Shpeisman}},
  \bibinfo{person}{N. {Vasilache}}, {and} \bibinfo{person}{O. {Zinenko}}.}
  \bibinfo{year}{2021}\natexlab{}.
\newblock \showarticletitle{MLIR: Scaling Compiler Infrastructure for Domain
  Specific Computation}. In \bibinfo{booktitle}{\emph{2021 IEEE/ACM
  International Symposium on Code Generation and Optimization (CGO)}}.
  \bibinfo{pages}{2--14}.
\newblock
\urldef\tempurl%
\url{https://doi.org/10.1109/CGO51591.2021.9370308}
\showDOI{\tempurl}


\bibitem[\protect\citeauthoryear{Li, Liu, Liu, Sun, You, Yang, Luan, and
  Qian}{Li et~al\mbox{.}}{2020}]%
        {li2020deep}
\bibfield{author}{\bibinfo{person}{Mingzhen Li}, \bibinfo{person}{Yi Liu},
  \bibinfo{person}{Xiaoyan Liu}, \bibinfo{person}{Qingxiao Sun},
  \bibinfo{person}{Xin You}, \bibinfo{person}{Hailong Yang},
  \bibinfo{person}{Zhongzhi Luan}, {and} \bibinfo{person}{Depei Qian}.}
  \bibinfo{year}{2020}\natexlab{}.
\newblock \showarticletitle{The Deep Learning Compiler: A Comprehensive
  Survey}.
\newblock \bibinfo{journal}{\emph{arXiv preprint arXiv:2002.03794}}
  (\bibinfo{year}{2020}).
\newblock


\bibitem[\protect\citeauthoryear{Liu, Hu, Zhu, Matusik, and Sifakis}{Liu
  et~al\mbox{.}}{2018}]%
        {liu2018narrow}
\bibfield{author}{\bibinfo{person}{Haixiang Liu}, \bibinfo{person}{Yuanming
  Hu}, \bibinfo{person}{Bo Zhu}, \bibinfo{person}{Wojciech Matusik}, {and}
  \bibinfo{person}{Eftychios Sifakis}.} \bibinfo{year}{2018}\natexlab{}.
\newblock \showarticletitle{Narrow-band Topology Optimization on a Sparsely
  Populated Grid}.
\newblock \bibinfo{journal}{\emph{ACM Trans. Graph. (Proc. SIGGRAPH Asia)}}
  \bibinfo{volume}{37}, \bibinfo{number}{6} (\bibinfo{year}{2018}),
  \bibinfo{pages}{251:1--251:14}.
\newblock


\bibitem[\protect\citeauthoryear{Maydan, Amarasinghe, and Lam}{Maydan
  et~al\mbox{.}}{1993}]%
        {maydan1993array}
\bibfield{author}{\bibinfo{person}{Dror~E Maydan}, \bibinfo{person}{Saman~P
  Amarasinghe}, {and} \bibinfo{person}{Monica~S Lam}.}
  \bibinfo{year}{1993}\natexlab{}.
\newblock \showarticletitle{Array-data flow analysis and its use in array
  privatization}. In \bibinfo{booktitle}{\emph{Proceedings of the 20th ACM
  SIGPLAN-SIGACT symposium on Principles of programming languages}}.
  \bibinfo{pages}{2--15}.
\newblock


\bibitem[\protect\citeauthoryear{Museth}{Museth}{2013}]%
        {museth2013vdb}
\bibfield{author}{\bibinfo{person}{Ken Museth}.}
  \bibinfo{year}{2013}\natexlab{}.
\newblock \showarticletitle{VDB: High-resolution sparse volumes with dynamic
  topology}.
\newblock \bibinfo{journal}{\emph{ACM Trans. Graph.}} \bibinfo{volume}{32},
  \bibinfo{number}{3} (\bibinfo{year}{2013}), \bibinfo{pages}{27}.
\newblock


\bibitem[\protect\citeauthoryear{Museth, Lait, Johanson, Budsberg, Henderson,
  Alden, Cucka, Hill, and Pearce}{Museth et~al\mbox{.}}{2013}]%
        {museth2013openvdb}
\bibfield{author}{\bibinfo{person}{Ken Museth}, \bibinfo{person}{Jeff Lait},
  \bibinfo{person}{John Johanson}, \bibinfo{person}{Jeff Budsberg},
  \bibinfo{person}{Ron Henderson}, \bibinfo{person}{Mihai Alden},
  \bibinfo{person}{Peter Cucka}, \bibinfo{person}{David Hill}, {and}
  \bibinfo{person}{Andrew Pearce}.} \bibinfo{year}{2013}\natexlab{}.
\newblock \showarticletitle{OpenVDB: an open-source data structure and toolkit
  for high-resolution volumes}.
\newblock In \bibinfo{booktitle}{\emph{Acm siggraph 2013 courses}}.
  \bibinfo{pages}{1--1}.
\newblock


\bibitem[\protect\citeauthoryear{Nielsen and Bridson}{Nielsen and
  Bridson}{2016}]%
        {nielsen2016spatially}
\bibfield{author}{\bibinfo{person}{Michael~B Nielsen} {and}
  \bibinfo{person}{Robert Bridson}.} \bibinfo{year}{2016}\natexlab{}.
\newblock \showarticletitle{Spatially adaptive FLIP fluid simulations in
  bifrost}. In \bibinfo{booktitle}{\emph{ACM SIGGRAPH 2016 Talks}}. ACM,
  \bibinfo{pages}{41}.
\newblock


\bibitem[\protect\citeauthoryear{Paszke, Gross, Massa, Lerer, Bradbury, Chanan,
  Killeen, Lin, Gimelshein, Antiga, Desmaison, Kopf, Yang, DeVito, Raison,
  Tejani, Chilamkurthy, Steiner, Fang, Bai, and Chintala}{Paszke
  et~al\mbox{.}}{2019}]%
        {NEURIPS2019_9015}
\bibfield{author}{\bibinfo{person}{Adam Paszke}, \bibinfo{person}{Sam Gross},
  \bibinfo{person}{Francisco Massa}, \bibinfo{person}{Adam Lerer},
  \bibinfo{person}{James Bradbury}, \bibinfo{person}{Gregory Chanan},
  \bibinfo{person}{Trevor Killeen}, \bibinfo{person}{Zeming Lin},
  \bibinfo{person}{Natalia Gimelshein}, \bibinfo{person}{Luca Antiga},
  \bibinfo{person}{Alban Desmaison}, \bibinfo{person}{Andreas Kopf},
  \bibinfo{person}{Edward Yang}, \bibinfo{person}{Zachary DeVito},
  \bibinfo{person}{Martin Raison}, \bibinfo{person}{Alykhan Tejani},
  \bibinfo{person}{Sasank Chilamkurthy}, \bibinfo{person}{Benoit Steiner},
  \bibinfo{person}{Lu Fang}, \bibinfo{person}{Junjie Bai}, {and}
  \bibinfo{person}{Soumith Chintala}.} \bibinfo{year}{2019}\natexlab{}.
\newblock \showarticletitle{PyTorch: An Imperative Style, High-Performance Deep
  Learning Library}.
\newblock In \bibinfo{booktitle}{\emph{Advances in Neural Information
  Processing Systems 32}}, \bibfield{editor}{\bibinfo{person}{H.~Wallach},
  \bibinfo{person}{H.~Larochelle}, \bibinfo{person}{A.~Beygelzimer},
  \bibinfo{person}{F.~d\textquotesingle Alch\'{e}-Buc},
  \bibinfo{person}{E.~Fox}, {and} \bibinfo{person}{R.~Garnett}} (Eds.).
  \bibinfo{publisher}{Curran Associates, Inc.}, \bibinfo{pages}{8024--8035}.
\newblock
\urldef\tempurl%
\url{http://papers.neurips.cc/paper/9015-pytorch-an-imperative-style-high-performance-deep-learning-library.pdf}
\showURL{%
\tempurl}


\bibitem[\protect\citeauthoryear{Qiao, Reiche, Hannig, and Teich}{Qiao
  et~al\mbox{.}}{2018}]%
        {qiao2018automatic}
\bibfield{author}{\bibinfo{person}{Bo Qiao}, \bibinfo{person}{Oliver Reiche},
  \bibinfo{person}{Frank Hannig}, {and} \bibinfo{person}{J{\"u}rgen Teich}.}
  \bibinfo{year}{2018}\natexlab{}.
\newblock \showarticletitle{Automatic kernel fusion for image processing DSLs}.
  In \bibinfo{booktitle}{\emph{Proceedings of the 21st International Workshop
  on Software and Compilers for Embedded Systems}}. \bibinfo{pages}{76--85}.
\newblock


\bibitem[\protect\citeauthoryear{Rotem, Fix, Abdulrasool, Catron, Deng,
  Dzhabarov, Gibson, Hegeman, Lele, Levenstein, et~al\mbox{.}}{Rotem
  et~al\mbox{.}}{2018}]%
        {rotem2018glow}
\bibfield{author}{\bibinfo{person}{Nadav Rotem}, \bibinfo{person}{Jordan Fix},
  \bibinfo{person}{Saleem Abdulrasool}, \bibinfo{person}{Garret Catron},
  \bibinfo{person}{Summer Deng}, \bibinfo{person}{Roman Dzhabarov},
  \bibinfo{person}{Nick Gibson}, \bibinfo{person}{James Hegeman},
  \bibinfo{person}{Meghan Lele}, \bibinfo{person}{Roman Levenstein},
  {et~al\mbox{.}}} \bibinfo{year}{2018}\natexlab{}.
\newblock \showarticletitle{Glow: Graph lowering compiler techniques for neural
  networks}.
\newblock \bibinfo{journal}{\emph{arXiv preprint arXiv:1805.00907}}
  (\bibinfo{year}{2018}).
\newblock


\bibitem[\protect\citeauthoryear{Selle, Fedkiw, Kim, Liu, and Rossignac}{Selle
  et~al\mbox{.}}{2008}]%
        {selle2008unconditionally}
\bibfield{author}{\bibinfo{person}{Andrew Selle}, \bibinfo{person}{Ronald
  Fedkiw}, \bibinfo{person}{Byungmoon Kim}, \bibinfo{person}{Yingjie Liu},
  {and} \bibinfo{person}{Jarek Rossignac}.} \bibinfo{year}{2008}\natexlab{}.
\newblock \showarticletitle{An unconditionally stable MacCormack method}.
\newblock \bibinfo{journal}{\emph{Journal of Scientific Computing}}
  \bibinfo{volume}{35}, \bibinfo{number}{2-3} (\bibinfo{year}{2008}),
  \bibinfo{pages}{350--371}.
\newblock


\bibitem[\protect\citeauthoryear{Setaluri, Aanjaneya, Bauer, and
  Sifakis}{Setaluri et~al\mbox{.}}{2014}]%
        {setaluri2014spgrid}
\bibfield{author}{\bibinfo{person}{Rajsekhar Setaluri}, \bibinfo{person}{Mridul
  Aanjaneya}, \bibinfo{person}{Sean Bauer}, {and} \bibinfo{person}{Eftychios
  Sifakis}.} \bibinfo{year}{2014}\natexlab{}.
\newblock \showarticletitle{{SPGrid}: A sparse paged grid structure applied to
  adaptive smoke simulation}.
\newblock \bibinfo{journal}{\emph{ACM Trans. Graph. (Proc. SIGGRAPH Asia)}}
  \bibinfo{volume}{33}, \bibinfo{number}{6} (\bibinfo{year}{2014}),
  \bibinfo{pages}{205}.
\newblock


\bibitem[\protect\citeauthoryear{Team, Al-Rfou, Alain, Almahairi, Angermueller,
  Bahdanau, Ballas, Bastien, Bayer, Belikov, et~al\mbox{.}}{Team
  et~al\mbox{.}}{2016}]%
        {team2016theano}
\bibfield{author}{\bibinfo{person}{The Theano~Development Team},
  \bibinfo{person}{Rami Al-Rfou}, \bibinfo{person}{Guillaume Alain},
  \bibinfo{person}{Amjad Almahairi}, \bibinfo{person}{Christof Angermueller},
  \bibinfo{person}{Dzmitry Bahdanau}, \bibinfo{person}{Nicolas Ballas},
  \bibinfo{person}{Fr{\'e}d{\'e}ric Bastien}, \bibinfo{person}{Justin Bayer},
  \bibinfo{person}{Anatoly Belikov}, {et~al\mbox{.}}}
  \bibinfo{year}{2016}\natexlab{}.
\newblock \showarticletitle{Theano: A Python framework for fast computation of
  mathematical expressions}.
\newblock \bibinfo{journal}{\emph{arXiv preprint arXiv:1605.02688}}
  (\bibinfo{year}{2016}).
\newblock


\bibitem[\protect\citeauthoryear{Vasilache, Zinenko, Theodoridis, Goyal,
  DeVito, Moses, Verdoolaege, Adams, and Cohen}{Vasilache
  et~al\mbox{.}}{2018}]%
        {vasilache2018tensor}
\bibfield{author}{\bibinfo{person}{Nicolas Vasilache},
  \bibinfo{person}{Oleksandr Zinenko}, \bibinfo{person}{Theodoros Theodoridis},
  \bibinfo{person}{Priya Goyal}, \bibinfo{person}{Zachary DeVito},
  \bibinfo{person}{William~S Moses}, \bibinfo{person}{Sven Verdoolaege},
  \bibinfo{person}{Andrew Adams}, {and} \bibinfo{person}{Albert Cohen}.}
  \bibinfo{year}{2018}\natexlab{}.
\newblock \showarticletitle{Tensor comprehensions: Framework-agnostic
  high-performance machine learning abstractions}.
\newblock \bibinfo{journal}{\emph{arXiv preprint arXiv:1802.04730}}
  (\bibinfo{year}{2018}).
\newblock


\bibitem[\protect\citeauthoryear{Wang, Qiu, Slattery, Fang, Li, Zhu, Zhu, Tang,
  Manocha, and Jiang}{Wang et~al\mbox{.}}{2020}]%
        {wang2020massively}
\bibfield{author}{\bibinfo{person}{Xinlei Wang}, \bibinfo{person}{Yuxing Qiu},
  \bibinfo{person}{Stuart~R Slattery}, \bibinfo{person}{Yu Fang},
  \bibinfo{person}{Minchen Li}, \bibinfo{person}{Song-Chun Zhu},
  \bibinfo{person}{Yixin Zhu}, \bibinfo{person}{Min Tang},
  \bibinfo{person}{Dinesh Manocha}, {and} \bibinfo{person}{Chenfanfu Jiang}.}
  \bibinfo{year}{2020}\natexlab{}.
\newblock \showarticletitle{A massively parallel and scalable multi-gpu
  material point method}.
\newblock \bibinfo{journal}{\emph{ACM Transactions on Graphics (TOG)}}
  \bibinfo{volume}{39}, \bibinfo{number}{4} (\bibinfo{year}{2020}),
  \bibinfo{pages}{30--1}.
\newblock


\bibitem[\protect\citeauthoryear{Wu, Truong, Yuksel, and Hoetzlein}{Wu
  et~al\mbox{.}}{2018}]%
        {wu2018fast}
\bibfield{author}{\bibinfo{person}{Kui Wu}, \bibinfo{person}{Nghia Truong},
  \bibinfo{person}{Cem Yuksel}, {and} \bibinfo{person}{Rama Hoetzlein}.}
  \bibinfo{year}{2018}\natexlab{}.
\newblock \showarticletitle{Fast fluid simulations with sparse volumes on the
  {GPU}}. In \bibinfo{booktitle}{\emph{Computer Graphics Forum (Proc.
  Eurographics)}}, Vol.~\bibinfo{volume}{37}. Wiley Online Library,
  \bibinfo{pages}{157--167}.
\newblock


\bibitem[\protect\citeauthoryear{Zerrell and Bruestle}{Zerrell and
  Bruestle}{2019}]%
        {zerrell2019stripe}
\bibfield{author}{\bibinfo{person}{Tim Zerrell} {and} \bibinfo{person}{Jeremy
  Bruestle}.} \bibinfo{year}{2019}\natexlab{}.
\newblock \showarticletitle{Stripe: Tensor compilation via the nested
  polyhedral model}.
\newblock \bibinfo{journal}{\emph{arXiv preprint arXiv:1903.06498}}
  (\bibinfo{year}{2019}).
\newblock


\end{thebibliography}

\clearpage


\section*{Supplementary material}

# AsyncTaichi: On-the-fly Inter-kernel Optimizations for Imperative and Spatially Sparse Programming (Supplemental Document)

YUANMING HU*, Taichi Graphics & MIT CSAIL
MINGKUAN XU*, Taichi Graphics & Tsinghua University
YE KUANG, Taichi Graphics
FRÉDO DURAND, MIT CSAIL

## 1 MICROBENCHMARK CASES

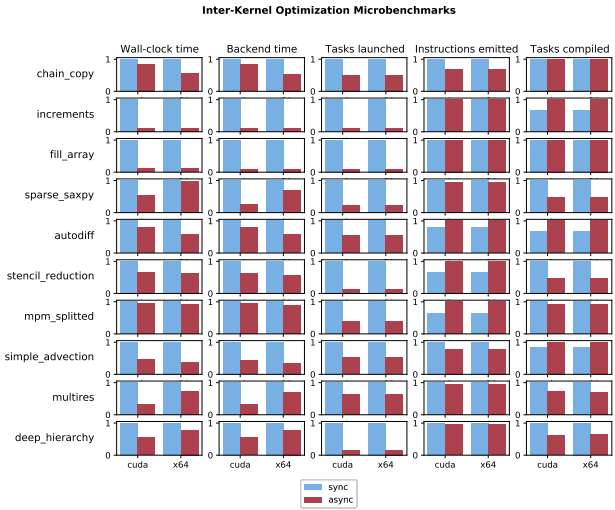

Fig. 1. Microbenchmarks results. "Sync" refers to the reference system [Hu et al. 2019]. "Async" refers to our system with asynchronous execution and inter-kernel optimizations.

Here we describe the cases in the microbenchmarks.

The case chain_copy contains 2 kernels `y[i] = x[i] + 1` and `z[i] = y[i] + 4`, like Fig.8 in the main paper. They are fused in asynchronous mode.

increments contains 10 inc() kernels (`x[i] += 1`) in Fig.8 in the main paper.

fill_array contains 10 kernels, all filling a 1-D dense array with the same constant value. With task fusion, only 1 task is launched in these cases instead of 10 tasks. The running time is nearly 10x faster.

sparse_saxpy contains some kernels performing saxpy (Scalar Alpha X Plus Y) operations among sparse tensors. The performance boost of execution time comes from the elimination of list generation and task fusion. Sometimes the wall-clock time is slower than the synchronous mode because of the overhead of the asynchronous engine.

*Both authors contributed equally to this work.

Authors' addresses: Yuanming Hu, Taichi Graphics & MIT CSAIL, yuanming@taichi.graphics; Mingkuan Xu, Taichi Graphics & Tsinghua University, xmk17@mails.tsinghua.edu.cn; Ye Kuang, Taichi Graphics, yekuang@taichi.graphics; Frédo Durand, MIT CSAIL, fredo@mit.edu.

autodiff computes a loss function as reduction on an array and accumulates the gradients to another array 10 times. With dead store elimination, the forward tasks computing the loss function should be eliminated except for the last one, so the number of launched tasks reduces by roughly a half.

stencil_reduction performs stencil and reduce operations on a tensor. They are common operations in computer graphics.

mpm_splitted contains some substep() kernels in an MPM program [Hu et al. 2018].

simple_advection performs semi-Lagrangian advection 10 times. The performance boost comes from activation elimination and task fusion.

multires is a multi-resolution program downsampling in 4 levels.

deep_hierarchy contains 5 jitter() kernels `x[i] += x[i + 1]` when `i % 2 == 0`. The tasks are not fusible, but we can still get some performance boost by eliminating list generation tasks.

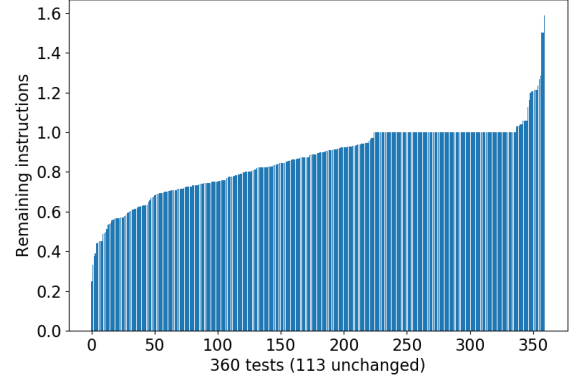

Fig. 2. Ratio of emitted instructions with/without data-flow optimization among unit tests. On most cases fewer instructions are emitted. The cases with more instructions do happen, because data-flow optimization indirectly triggers some other optimization passes that lead to high-performance but also more (and cheaper) instructions.

## 2 DETAILS IN TASK FUSION

*Fusing criteria.* We fuse tasks A and B when all the following criteria are satisfied:

(1) There is no path of length ≥ 2 between A and B in the SFG. If we fuse tasks A and B when there is a path A→C→B, C can neither depend on the fused task nor become its prerequisite, leading to a contradiction.

(2) Either both tasks are serial, or both tasks are parallel with the same loop ranges (if the tasks' type is range-for) or the same SNode (if the tasks' type is struct-for).
(3) If A→B are connected via a SNode in the SFG, both tasks need to access the SNode in an element-wise manner. Failing to do so may lead to data races. For example, if there is an edge A→B (corresponding to some SNode) in the SFG, we need every accesses to that SNode in both tasks are at the same address, and that address is unique per iteration of the loop. When not every accesses on the SNode are on the same address, we may bump into a racing condition if we fuse the tasks, for example,

```
for i in x:
    x[i] = 1
for i in x:
    x[i + 1] = 2
```

The access to `x[i + 1]` at the `i`-th iteration is racing with `x[i]` at the `(i + 1)`-th iteration when fused. If we access some address which is not unique per iteration of the loop, we may also bump into a racing condition if we fuse the tasks. Example:

```
for i in x:
    y[0] = 0
for i in x:
    y[0] += x[i]
```

These two tasks compute the sum of `x` and store it in `y[0]`, but if we fuse the tasks, they will store only one element in `x` to `y[0]` because the access of `y[0]` is not unique per iteration of the loop.
(4) When there are multiple tasks that can be fused together, we introduce a compilation option `async_max_fuse_per_task` with a default value 1, and greedily fuse each task no more than that times per iteration of the task fusion optimization pass. In this way, when there are $n$ tasks that can be fused together, we will perform the task fusion optimization for $O(\log(n))$ iterations, and avoid fusing $n$ tasks one by one into a huge task. This makes the intra-kernel optimizations after task fusion faster, and utilizes the IR bank better when there are duplicated tasks. See section 7.2 in the main paper for more details.

To find all fusible pairs of tasks efficiently, we compute the transitive closure of the SFG using bitsets. For pairs of tasks without edges, we group tasks by the tasks' type, loop range (if the type is range-for), or the SNode (if the type is struct-for). For each group, we use the transitive closure to find which pairs of tasks do not have any path to each other quickly; for each edge A → B in the SFG, we check if there is a task C such that A has a path to C and C has a path to B using the transitive closure. The two cases above cover the criterion "there is no path of length ≥ 2 between A and B". Then we apply the other criteria to find if the pair of tasks is fusible, and greedily fuse them.

## 3 INTRA-KERNEL DATA-FLOW OPTIMIZATIONS

We apply the traditional control-flow analysis to optimize within kernels. We build a control-flow graph along with the hierarchical IR, and perform analysis on the graph. With the help of control-flow analysis, we perform optimizations including store-to-load forwarding, dead store elimination, and identical load/store elimination. These optimizations motivate task fusion as it greatly simplifies fused tasks.

We also utilize control-flow analysis to help compute task meta information. Since stores to a SNode may only partially modify a value state, the resulting value state (which contains the modified and unmodified part) may need a read from the previous version of the value state. We use control-flow analysis to detect which SNodes do not need a read from the previous version of the value state.

Fig. 2 shows the effect of data-flow optimization on 360 Taichi test cases. Although these test cases are relatively simple, data-flow optimization still leads to 16% fewer instructions.

## 4 EVALUATION
### 4.1 MGPCG at different scales

To evaluate our system on problems of different scales, we scan the problem size and plot the wall-clock time and backend time in Fig. 3. Most of the optimizations come from eliminating fragmented kernels caused by list generation. As the problems scale up, these kernels occupy relatively less time, hence we observe the performance boost to be more significant at lower resolutions. An ablation study of four inter-kernel optimization passes is listed in Table 5 in the main paper.

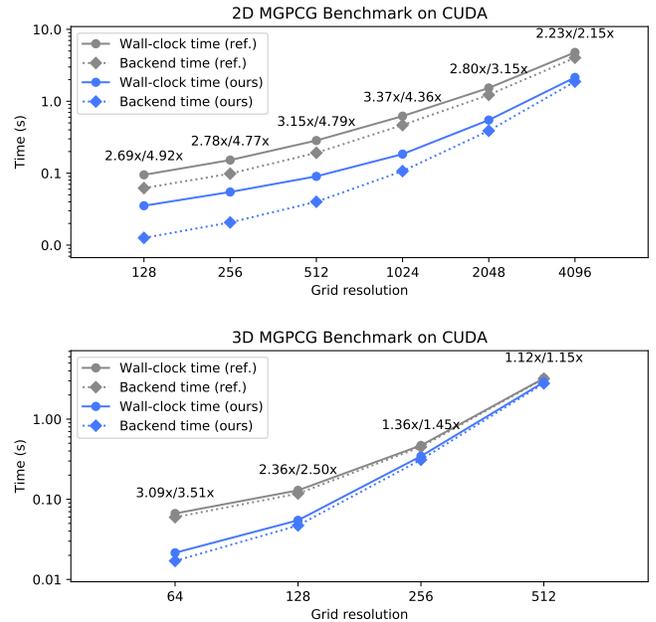

Fig. 3. MGPCG benchmarks on CUDA. The annotation numbers on each data point indicate the acceleration on *wall-clock* time and *backend* time. Most of the performance boost in this example comes from list generation removal. Since list generation takes a smaller portion of total computation when resolution increases, the gap between ours and reference performance shrinks as resolution goes up. [Reproduce: **python3 plot_mgpcg.py -d [2/3]**]

## 4.2 MacCormack advection benchmark setup

This benchmark is carried out on a 3D unit grid of $256^3$ cells. We advect three scalar physical fields (temperature, color, and density) over a sparsely populated vector field defined over a tube domain with inner radius 0.32 and outer radius 0.45. The statistics of the first five frames are discarded to exclude the effects of the JIT compilation.

## 4.3 MLS-MPM benchmark setup

As shown in Fig. 4, we seed a uniform cube of 16,777,216 ($256^3$, on GPU) or 2,097,152 ($128^3$, on x64 CPUs) fixed corotated particles into a grid of size $1 \times 1 \times 1$, with the distance between each two adjacent particles 1/512. Each particle is a cube of length 1/512, with $\rho = 10^3, E = 2 \times 10^4, \nu = 0.4$. The gravity is $g = 9.8$. Each frame consists of 400 substeps, with time step size $\Delta t = 10^{-4}$s. We benchmark the time of the second frame (i.e., the 401st to the 800th substeps). The ground is set to be at $z = 1/32$, and to make the cube touch the ground in the second frame, we set the initial $z$-coordinate of the center of the bottom-most particles to be 49/1024. The $x$ and $y$ coordinates of the center of the cube are set to be the center of the grid.

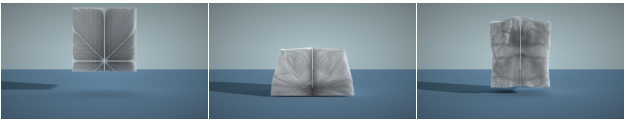

Fig. 4. The falling cube in our MLS-MPM benchmark. The cube's initial position is higher than the one we used in the benchmark for better visualization.

## 4.4 Compiler hints in MLS-MPM

In MLS-MPM, the G2P task writes the values in the `particle` SNode, and the P2G task reads the values in the `particle` SNode. Since both tasks are parallel, it is challenging to infer that there will be no racing conditions if we fuse the two tasks together. Therefore, we hint our compiler with a simple statement (in `mpm_solver.py`):

```
p = ti.loop_unique(p, covers=self.particle)
```

where `p` is the particle ID in one iteration of the loop in the G2P task. This statements means that `p` is unique per iteration of the loop, and the set of its value among all iterations of the loop covers all active indices in the `particle` SNode (so the values stored in the `pid` (particle ID) field is a permutation of all active indices in the `particle` SNode). With this hint, we can infer that there will be no racing conditions if we fuse the G2P and P2G tasks together since different iterations of these tasks access different addresses. We further analyze that the `C` field (affine velocity around the particle), as a component in the `particle` SNode, is fully written in the G2P task (or the fused G2P2G [Wang et al. 2020] task), so we do not need to write to this field until the last G2P task.

## 5 KEY SOURCE CODE FILES

We partially reuse the existing Taichi system, with key modifications listed below:

- The SFG data structure and some optimization passes are implemented in `taichi/program/state_flow_graph.[h/cpp]`;
- The asynchronous execution engine is implemented in `taichi/program/async_engine.[h/cpp]`;
- The IR bank and some of the optimization passes are implemented in `taichi/program/ir_bank.[h/cpp]`.

Other low-level implementation details are scattered in the whole Taichi codebase.

## REFERENCES

Yuanming Hu, Yu Fang, Ziheng Ge, Ziyin Qu, Yixin Zhu, Andre Pradhana, and Chenfanfu Jiang. 2018. A moving least squares material point method with displacement discontinuity and two-way rigid body coupling. *ACM Trans. Graph. (Proc. SIGGRAPH Asia)* 37, 4 (2018), 150.

Yuanming Hu, Tzu-Mao Li, Luke Anderson, Jonathan Ragan-Kelley, and Frédo Durand. 2019. Taichi: a language for high-performance computation on spatially sparse data structures. *ACM Transactions on Graphics (TOG)* 38, 6 (2019), 201.

Xinlei Wang, Yuxing Qiu, Stuart R Slattery, Yu Fang, Minchen Li, Song-Chun Zhu, Yixin Zhu, Min Tang, Dinesh Manocha, and Chenfanfu Jiang. 2020. A massively parallel and scalable multi-gpu material point method. *ACM Transactions on Graphics (TOG)* 39, 4 (2020), 30–1.



\end{document}